\documentclass[structabstract]{aa}  
\usepackage{graphicx}
\usepackage{txfonts}
\usepackage{natbib}
\usepackage{longtable}
\begin{document}
\title{Fragmentation and disk formation during high-mass star formation}

\subtitle{The IRAM NOEMA (Northern Extended Millimeter Array) large program CORE}

   \author{H.~Beuther
          \inst{1}
          \and
          J.C.~Mottram
          \inst{1}
          \and
          A.~Ahmadi
          \inst{1}
           \and
          F.~Bosco
          \inst{1}
         \and
          H.~Linz
          \inst{1}
          \and
          Th.~Henning
          \inst{1}
          \and
          P.~Klaassen
          \inst{2}
          \and
          J.M. Winters
          \inst{3}
          \and
           L.T.~Maud
          \inst{4}
          \and
          R.~Kuiper
          \inst{5}
          \and
          D.~Semenov
          \inst{1}
          \and
          C.~Gieser
          \inst{1}
          \and
          T.~Peters
          \inst{6}
          \and
           J.S.~Urquhart
          \inst{7}
          \and
           R.~Pudritz
          \inst{8}
          \and
           S.E.~Ragan
          \inst{9}
          \and
           S.~Feng
          \inst{10}
          \and
           E.~Keto
          \inst{11}
          \and
           S.~Leurini
          \inst{12}
          \and
           R.~Cesaroni
          \inst{13}
          \and
           M.~Beltran
          \inst{13}
          \and
           A.~Palau
          \inst{14}
          \and
           \'A. S\'anchez-Monge
          \inst{15}
          \and
           R.~Galvan-Madrid
          \inst{14}
          \and
           Q.~Zhang
          \inst{11}
          \and
           P.~Schilke
          \inst{15}
          \and
           F.~Wyrowski
          \inst{16}
          \and
           K.G.~Johnston
          \inst{17}
         \and
           S.N.~Longmore
          \inst{18}
         \and
           S.~Lumsden
          \inst{17}
         \and
           M.~Hoare
          \inst{17}
          \and
           K.M.~Menten
          \inst{16}
          \and
           T.~Csengeri
          \inst{16}
}
   \institute{$^1$ Max Planck Institute for Astronomy, K\"onigstuhl 17,
              69117 Heidelberg, Germany, \email{name@mpia.de}\\
              $^2$ UK Astronomy Technology Centre, Royal Observatory Edinburgh, Blackford Hill, Edinburgh EH9 3HJ, UK\\
              $^3$ IRAM, 300 rue de la Piscine, Domaine Universitaire de Grenoble, 38406 St.-Martin-d’H\`eres, France\\
              $^{4}$ Leiden Observatory, Leiden University, PO Box 9513, NL-2300 RA Leiden, the Netherlands\\
              $^5$ Institute of Astronomy and Astrophysics, University of T\"ubingen, Auf der Morgenstelle 10, 72076, T\"ubingen, Germany\\
              $^6$ Max-Planck-Institut f\"{u}r Astrophysik, Karl-Schwarzschild-Str. 1, D-85748 Garching, Germany\\
              $^7$ Centre for Astrophysics and Planetary Science, University of Kent, Canterbury, CT2 7NH, UK\\
              $^8$ Department of Physics and Astronomy, McMaster University, 1280 Main St. W, Hamilton, ON L8S 4M1, Canada\\
              $^9$ School of Physics and Astronomy, Cardiff University, Cardiff CF24 3AA, UK\\
              $^{10}$ Max Planck Institut for Extraterrestrische Physik, Giessenbachstrasse 1, 85748 Garching, Germany\\
              $^{11}$ Harvard-Smithsonian Center for Astrophysics, 160 Garden St, Cambridge, MA 02420, USA\\
              $^{12}$ INAF - Osservatorio Astronomico di Cagliari, via della Scienza 5, 09047, Selargius (CA), Italy\\
              $^{13}$ INAF, Osservatorio Astrofisico di Arcetri, Largo E. Fermi 5, I-50125 Firenze, Italy\\
              $^{14}$ Instituto de Radioastronomıa y Astrofısica, Universidad Nacional Autonoma de Mexico, 58090 Morelia, Michoacan, Mexico\\
              $^{15}$  I. Physikalisches Institut, Universit\"at zu K\"oln, Z\"ulpicher Str. 77, D-50937, K\"oln, Germany\\
              $^{16}$ Max Planck Institut for Radioastronomie, Auf dem H\"ugel 69, 53121 Bonn, Germany\\
              $^{17}$ School of Physics \& Astronomy, E.C. Stoner Building, The University of Leeds, Leeds LS2 9JT, UK\\
              $^{18}$ Astrophysics Research Institute, Liverpool John Moores University, 146 Brownlow Hill, Liverpool L3 5RF, UK
}

   \date{Version of \today}

\abstract
{High-mass stars form in clusters, but neither the early fragmentation
  processes nor the detailed physical processes leading to the most
  massive stars are well understood.}
{We aim to understand the fragmentation as well as the disk formation,
  outflow generation and chemical processes during high-mass star
  formation on spatial scales of individual cores.}
{Using the IRAM Northern Extended Millimeter Array (NOEMA) in
  combination with the 30\,m telescope, we have observed in the IRAM
  large program CORE the 1.37\,mm continuum and spectral line emission
  at high angular resolution ($\sim$0.4$''$) for a sample of 20
  well-known high-mass star-forming regions with distances below
  5.5\,kpc and luminosities larger than $10^4$\,L$_{\odot}$.}
{We present the overall survey scope, the selected sample, the
  observational setup and the main goals of CORE. Scientifically, we
  concentrate on the mm continuum emission on scales on the order of
  1000\,AU. We detect strong mm continuum emission from all regions,
  mostly due to the emission from cold dust. The fragmentation
  properties of the sample are diverse. We see extremes where some
  regions are dominated by a single high-mass core whereas others
  fragment into as many as 20 cores. A minimum-spanning-tree analysis
  finds fragmentation at scales on the order of the thermal Jeans
  length or smaller suggesting that turbulent fragmentation is less
  important than thermal gravitational fragmentation. The diversity of
  highly fragmented versus singular regions can be explained by
  varying initial density structures and/or different initial magnetic
  field strengths. }
{A large sample of high-mass star-forming regions at high spatial
  resolution allows us to study the fragmentation properties of
  young cluster-forming regions. The smallest observed separations
  between cores are found around the angular resolution limit which
  indicates that further fragmentation likely takes place on even
  smaller spatial scales. The CORE project with its numerous spectral
  line detections will address a diverse set of important physical and
  chemical questions in the field of high-mass star formation.}
\keywords{Stars: formation -- Stars: massive -- Stars: individual:
  IRAS23151, IRAS23033, AFGL2591, G75.78, S87IRS1, S106, IRAS21078,
  G100.38, G084.95, G094.60, CepA, NGC7538IRS9, W3(H$_2$O)/W3(OH),
  W3IRS4, G108.76, IRAS23385, G138.30, G139.91, NGC7538IRS1, NGC7538S
  -- Stars: rotation -- Instrumentation: interferometers}

\titlerunning{CORE: Fragmentation and disk formation}

\maketitle

\section{Introduction}
\label{intro}

The central questions in high-mass star formation research focus on
the fragmentation properties of the initial gas clumps that ultimately
result in the final clusters, and the disk formation and accretion
processes around the most massive young stars within these
clusters. Furthermore, related processes such as the overall gas
inflow, energetic molecular outflows and the rich chemistry in these
environments are still not comprehensively understood. For detailed
discussions about these topics we refer to, e.g.,
\citet{beuther2006b,zinnecker2007,tan2014,frank2014,reipurth2014,li2014,beltran2016,motte2017}.

Since high-mass star formation proceeds in a clustered mode at
distances mostly of several kpc, high spatial resolution is mandatory
to resolve the different physical processes. In addition, much of the
future evolution is likely set during the earliest and still cold
molecular phase, so observations at mm wavelengths are the path to
follow. Most high-resolution investigations in the last decade
targeted individual regions, but they did not address the topics of
fragmentation, disk formation and accretion in a statistical sense. A
notable exception is the fragmentation study by
\citet{palau2013,palau2014} who compiled a literature sample comprised
largely of intermediate- rather than high-mass star-forming
regions. However, fragmentation needs to be further studied in diverse
samples, recovering larger spatial scales, and including regions of
higher masses, in order to test how fragmentation behaves over a broad
range of properties in high-mass star-forming regions.

To overcome these limitations, we conducted an IRAM Northern Extended
Millimeter Array (NOEMA) large program named CORE: ``Fragmentation and
disk formation in high-mass star formation''. This program covered a
sample of 20 high-mass star-forming regions at high angular resolution
($\sim 0.3''-0.4''$ corresponding to roughly 1000\,AU at a typical
3\,kpc distance) in the 1.3\,mm band in the continuum and spectral
line emission. The main scientific questions to be addressed with this
survey are: (a) What are the fragmentation properties of high-mass
star-forming regions during the early evolutionary stages of cluster
formation? (b) Can we identify genuine high-mass accretion disks, and
if yes, what are their properties? Are rotating structures large
gravitationally (un)stable toroids and/or do embedded Keplerian
entities exist? Or are the latter embedded in the former? (c) How is
the gas accumulated into the central cores and what are the
larger-scale gas accretion flow and infall properties? Are the
high-density cores mainly isolated objects or continuously fed by
large-scale accretion flows/global gravitational collapse? (d) What
are the properties of the energetic outflows and how do they relate to
the underlying accretion disks? (e) What are the chemical properties
of distinct sub-structures within high-mass star-forming regions?

\begin{figure}[htb]
  \includegraphics[width=0.49\textwidth]{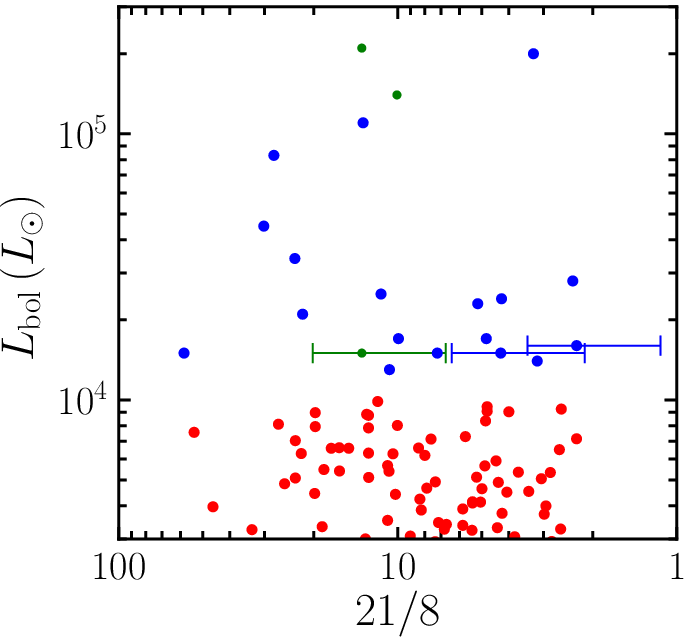}
  \caption{Sample selection plot where the luminosity (in units of
    L$_{\odot}$) is plotted against the MSX 21/8\,$\mu$m
    color. Horizontal bars mark uncertainties in the color. While the
    blue sources fulfill our selection criteria, the red ones are
    below our luminosity cut of $10^4$\,L$_{\odot}$. Green sources are
    those for which high-resolution mm data already exist and which
    were therefore excluded from the observations.}
\label{samplefig} 
\end{figure} 

Regarding cluster formation and the early fragmentation processes, it
is well established that high-mass stars typically form in a clustered
mode with a high degree of multiplicity (e.g.,
\citealt{zinnecker2007,bonnell2006,bressert2010,peters2010b,chini2012,peter2012,krumholz2014,reipurth2014}). Furthermore,
the dynamical interactions between cluster members may even dominate
their evolution (e.g.,
\citealt{gomez2005,sana2012}). High-spatial-resolution studies over
the last decades have shown that most massive gas clumps do not remain
single entities but fragment into multiple objects. However, the
degree of fragmentation varies between regions (e.g.,
\citealt{zhang2009,bontemps2010,pillai2011,wang2011,rodon2012,beuther2012c,palau2013,wang2014,csengeri2017,cesaroni2017}). The
previous data indicate that high-mass monolithic condensations may be
rare, but they could nevertheless exist (e.g.,
\citealt{bontemps2010,csengeri2017,sanchez-monge2017}). Going to
sub-arcsecond resolution, most regions indeed fragment, but exceptions
exist: For example, our recent investigations with the Plateau de Bure
Interferometer (PdBI, now renamed to NOEMA) of the famous high-mass
star-forming regions NGC7538IRS1 and NGC7538S revealed that NGC7538S
has fragmented into several sub-sources at $\sim 0.3''$ resolution
whereas at the same spatial resolution the central core of NGC7538IRS1
remains a single compact source (\citealt{beuther2012c}; see also
\citealt{qiu2011} for more extended cores in the environment). At an
even higher angular resolution of $\leq 0.2''$ or spatial scales below
1000\,AU, \citet{beuther2013b} found that even the innermost structure
of NGC7538IRS1 starts to fragment. This implies that the scales of
fragmentation do vary from region to region. Other fragmentation
studies do not entirely agree on the physical processes responsible
for driving the fragmentation. For example, the infrared dark cloud
study by \citet{wang2014} indicates that turbulence may be needed to
explain the large fragment masses. Similarly, \citet{pillai2011} argue
for two young pre-protocluster regions that turbulent Jeans
fragmentation can explain their data. However, other studies like
those by \citet{palau2013,palau2014,palau2015} favor pure
gravitational fragmentation. Similar results are also indicated in a
recent ALMA study towards a number of hypercompact H{\sc ii} regions
\citep{klaassen2018}. In addition to the thermal and turbulent gas
properties, theoretical as well as observational investigations
indicate the importance of the magnetic field for the fragmentation
processes during (high-mass) star formation
\citep{commercon2011,peters2011,tan2013,fontani2016}. Furthermore,
radiation feedback from forming protostars is also capable of reducing
the fragmentation of the high-mass star-forming region (e.g.,
\citealt{krumholz2007b}).

\begin{figure*}[htb]
  \includegraphics[width=0.99\textwidth]{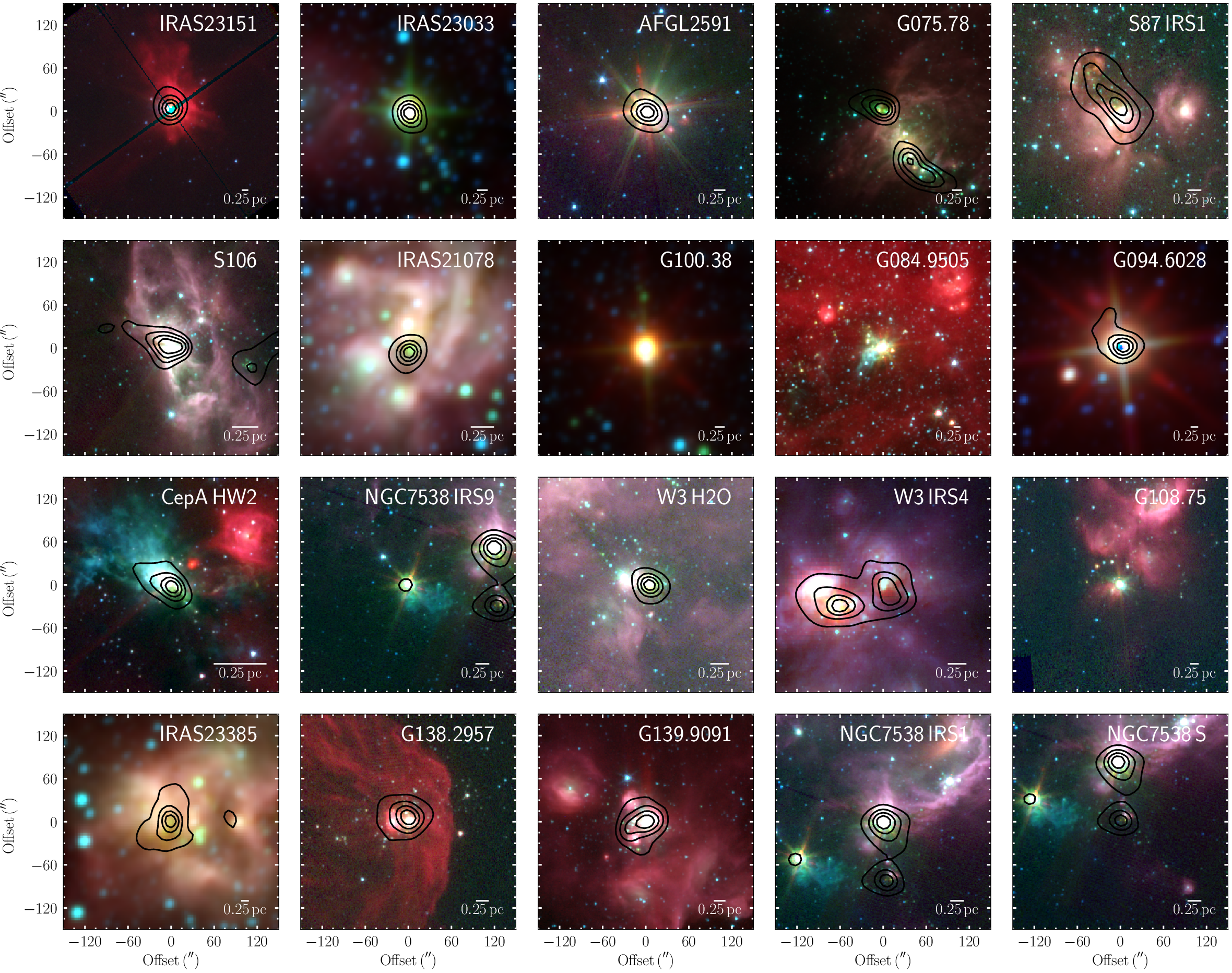}
  \caption{Large-scale overview images for the whole CORE sample. The
    color-scale show 3-color images with blue, green and red from
    Spitzer 3.6, 4.5 and 8.0\,$\mu$m for all sources except
    IRAS\,23033, IRAS21078, G100, G094 and IRAS\,23385 for which WISE
    3.4, 4.6 and 12\,$\mu$m data are presented. Furthermore, W3IRS4
    uses Spitzer 3.6, 4.5\,$\mu$m and MSX 8\,$\mu$m. The contours show
    SCUBA 850\,$\mu$m continuum data (\citealt{difrancesco2007};
    contour levels 20, 40, 60, 80\% of the peak emission) for all
    sources except G100, G084 and G108 where these data do not exist.}
\label{overview} 
\end{figure*} 

It is important to keep in mind that fragmentation occurs on all
scales, from large-scale molecular clouds down to the fragmentation of
disks (e.g., \citealt{dobbs2014,andre2014,kratter2016}). Different
fragmentation processes may dominate on different spatial scales. In
the continuum study presented here, we are concentrating on the
fragmentation of pc-scale clumps into cores with sizes of typically
several thousand AU. Smaller-scale disk fragmentation will also be
addressed by the CORE program (see section \ref{strategy}) through the
spectral line analysis of high-mass accretion disk candidates (e.g.,
Ahmadi et al., subm.).

The previous investigations of NGC7538IRS1 and NGC7538S
\citep{beuther2012c,beuther2013b,feng2016c} can be considered as a
pilot study for the CORE survey presented here. With an overall sample
of 20 high-mass star-forming regions (see sample selection below)
observed at uniform angular resolution ($\sim 0.3''-0.4''$) in the
1.3\,mm wavelength band with NOEMA, we can investigate how (un)typical
such fragmentation properties on core scales are. Fragmentation
signatures to be investigated are, for example, the fragment mass,
size and separation distributions, and how they relate to basic
underlying physical processes.

In this paper, we present the sample selection, the general survey
strategy as well as the observational characteristics. The rest of the
paper will then focus on the continuum data and the fragmentation
properties of the sample. The other scientific aspects of this survey
will be presented in separate publications (e.g., Ahmadi et al.~subm.,
Mottram et al.~in prep., Bosco et al.~in prep.).

\section{Sample}
\label{sec_sample}

\begin{table*}[htb]
\begin{center}
\caption{CORE Sample (grouped in track-sharing pairs)}
\label{sample}
\begin{tabular}{lrrrrrrrrrrrrr}
  \hline
  \hline
  Source & R.A. & Dec. & $\varv_{\rm{lsr}}$ & $D$ & $L$ & $M^a$ & $L/M$ & $S_{8\mu m}$ & $S_{21\mu m}$ & IR- & a.f.$^e$ & Ref. \\
  & (J2000.0) & (J2000.0) & $\left(\frac{\rm{km}}{\rm{s}}\right)$ & (kpc) & ($10^4$L$_{\odot}$) & (M$_{\odot}$) & $\left(\frac{\rm{L}_{\odot}}{\rm{M}_{\odot}}\right)$ & (Jy) & (Jy) & bright & & \\
  \hline
  IRAS23151+5912  & 23:17:21.01 & +59:28:47.49 & -54.4 & 3.3 & 2.4 & 215$^b$ & 112 & 23.8 & 101.1 & + & b   & d1,l2 \\
  IRAS23033+5951  & 23:05:25.00 & +60:08:15.49 & -53.1 & 4.3 & 1.7 & 495 & 34 & 5.0  & 24.0  & --&a,b  & d2,l1 \\
  \hline
  AFGL2591        & 20:29:24.86 & +40:11:19.40 & -5.5 & 3.3 & 20.0 & 638 & 313 & 313.8& 1023.4& + & a,b & d3,l1 \\
  G75.78+0.34     & 20:21:44.03 & +37:26:37.70 & -0.5 & 3.8 & 11.0 & 549 & 200 & 3.5  & 46.4  & --& a,c & d4,l1 \\
  \hline
  S87 IRS1        & 19:46:20.14 & +24:35:29.00 & 22.0 & 2.2 & 2.5  & 1421& 18  & 19.6 & 225.1 & + & a   & d5,l1 \\
  S106            & 20:27:26.77 & +37:22:47.70 & -1.0 & 1.3 & 3.4  & 47  & 723 & 53.1 & 1240.9& + & a,b & d6,l2 \\
  \hline
  IRAS21078+5211  & 21:09:21.64 & +52:22:37.50 & -6.1 & 1.5 & 1.3  & 177 & 73  & 2.1  & 8.8   & --& a,b &  dl1 \\ 
  G100.3779-03.578& 22:16:10.35 & +52:21:34.70 & -37.6 &3.5 & 1.5  & 206$^d$ & & 12.9 & 92.7 & + & b    & d1,l2\\
  \hline
  G084.9505-00.691& 20:55:32.47 & +44:06:10.10 & -34.6 & 5.5 & 1.3 & 648$^c$ & 20 & 1.4  & 14.6 & + & b    & d2,l2\\
  G094.6028-01.797& 21:39:58.25 & +50:14:20.90 & -43.6 & 4.0 & 2.8 & 1525& 18 & 63.9 & 150.5& + & b,c  & d1,l2 \\
  \hline
  CepAHW2         & 22:56:17.98 & +62:01:49.50 & -10.0 & 0.7 & 1.5 & 40  & 375 & 4.6  & 271.7& --& a,b,c& d7,l1 \\
  NGC7538IRS9     & 23:14:01.68 & +61:27:19.10 & -57.0 & 2.7 & 2.3 & 214 & 107 & 38.1 & 197.0& + & b    & d7,l1 \\
  \hline
  W3(H$_2$O)      & 02:27:04.60 & +61:52:24.73 & -48.5 & 2.0 & 8.3 & 307 & 270 & 10.7 & 298.9& --& a,b,c& d8,l2 \\
  W3IRS4          & 02:25:31.22 & +62:06:21.00 & -42.8 & 2.0 & 4.5 & 481 & 93  &15.4 & 465.2& + & a,b  & d8,l1\\
  \hline
  G108.7575-00.986& 22:58:47.25 & +58:45:01.60 & -51.5 & 4.3 & 1.4 &6204$^d$ & & 6.9  & 21.9 & + & b,c  & d2,l3\\
  IRAS23385+6053  & 23:40:54.40 & +61:10:28.20 & -50.2 & 4.9 & 1.6 & 510 & 31 & 1.6  & 3.5  & --& b    & dl2 \\
  \hline
  G138.2957+01.555& 03:01:31.32 & +60:29:13.20 & -37.5 & 2.9 & 1.4 & 197 & 71 & 9.1  & 90.0 & + & a,b  & d2,l1 \\
  G139.9091+00.197& 03:07:24.52 & +58:30:48.30 & -40.5 & 3.2 & 1.1 & 349 & 32 & 12.9 & 282.2& + & a,b  & d2,l1 \\
  \hline
  Pilot study\\
  NGC7538IRS1     & 23:13:45.36 & +61:28:10.55 & -57.3 & 2.7 & 21.0 & 1570& 133 & 109.2&1468.6& + &a,b,c& d7,l1 \\
  NGC7538S        & 23:13:44.86 & +61:26:48.10 & -56.4 & 2.7 & 1.5  & 238 & 63  & 1.1  & 15.3 & --& b,c  & d7 \\ 
  \hline
  \hline
\end{tabular}
\end{center}
{\footnotesize ~\\
  $^a$ Masses are calculated mainly from the SCUBA 850\,$\mu$m fluxes by \citet{difrancesco2008}.\\
  $^b$ Based on 1.2\,mm continuum data from \citet{beuther2002a}\\
  $^c$ Based on 1.1\,mm continuum data from \citet{ginsburg2013}\\
  $^d$ Based on C$^{18}$O(3--2) data from \citet{maud2015b}; effective radii for G100 $\sim$0.34\,pc and for G108 $\sim$1.4\,pc\\
  $^e$ Associated features (a.f.): a: cm continuum; b: H$_2$O maser; c: CH$_3$OH maser\\
  References for distances and luminosities: d1: \citealt{choi2014},
  d2: \citealt{urquhart2011}, d3: \citealt{rygl2012}, d4:
  \citealt{ando2011}, d5: \citealt{xu2009}, d6: \citealt{xu2013}, d7:
  \citealt{moscadelli2009}, d8: \citealt{hachisuka2006,xu2006}, dl1: \citealt{molinari1996}, dl2: \citealt{molinari1998b}\\
  l1: RMS survey database (http://rms.leeds.ac.uk/cgi-bin/public/RMS\_DATABASE.cgi), using SED fitting from \citet{mottram2011} including Herschel fluxes and the latest distance determination\\
  l2: RMS survey database
  (http://rms.leeds.ac.uk/cgi-bin/public/RMS\_DATABASE.cgi), using SED
  fitting from \citet{mottram2011} updated to the latest distance
  determination\\
  l3: RMS survey database
  (http://rms.leeds.ac.uk/cgi-bin/public/RMS\_DATABASE.cgi), calculated
  from the MSX 21\,$\mu$m flux using the scaling relation derived by
  \citet{mottram2011} and updated to the latest distance
  determination.}
\end{table*}

Our sample of young high-mass star-forming regions was selected to
fulfill several criteria: (a) luminosities $>10^4$\,L$_{\odot}$
indicating that at least an 8\,M$_{\odot}$ star is forming, (b)
distance-limited to below 6\,kpc to ensure high linear resolution
($\sim$1000\,AU), (c) high-declination sources (decl.$>$24$^{\circ}$)
to obtain the best possible uv-coverage (implying that they are either
not at all or at most poorly accessible with the Atacama Large
Millimeter Array, ALMA). Furthermore, only sources with extensive
complementary high-spatial resolution observations at other
wavelengths were selected to better characterize their overall
properties. In this context, the sample is also part of a large
e-Merlin project led by Co-I Melvin Hoare to characterize the cm
continuum emission of the sample at an anticipated spatial resolution
of down to 30\,mas. The initial luminosity selection was based on
luminosity and color-color criteria. Figure~\ref{samplefig} presents
the corresponding luminosity-color plot. We use the luminosity-color
plot as a sample selection tool as the y- and x-axes act as proxies
for stellar mass and evolutionary stage, respectively. By the time
massive forming stars have reached $10^4$\,L$_{\odot}$ the luminosity
is determined primarily by the stellar mass as at this stage the
accretion luminosity only contributes a small fraction of the total
luminosity even at high accretion rates (e.g.,
\citealt{hosokawa2009,hosokawa2010,kuiper2013,klassen2016}). We also
expect over time that the IR colors will evolve from red to blue as
the envelope material is dispersed and/or accreted (e.g.,
\citealt{zhang2014c}).

Many sample sources are covered by the RMS survey (Red MSX sources,
\citealt{lumsden2013}), and a few additional prominent northern
hemisphere regions are included as well.  Our sample excludes the few
sources that fulfill these selection criteria but which already have
been observed at mm wavelengths with high angular resolution (e.g.,
W3IRS5, NGC7538IRS1/S, \citealt{rodon2008,beuther2012c}). The
resulting sample of 18 regions is complete within these described
selection criteria. Because NGC7538IRS1 and NGC7538S were observed in
an almost identical setup (only the compact D-array data were not
taken), they are considered as a pilot study and their results are
incorporated into the analysis of the CORE project. Table \ref{sample}
presents a summary of the main source characteristics, including their
local-standard-of-rest velocity $\varv_{\rm lsr}$, distance $D$,
luminosity $L$, mass $M$ (see also section \ref{mass_calc}), their 8
and 21\,$\mu$m fluxes, H$_2$O, CH$_3$OH maser and cm continuum
associations as well as references for the distances and
luminosities. Figure \ref{overview} shows a larger-scale overview of
the twenty regions with the near- to mid-infrared data shown in color
and the 850\,$\mu$m continuum single-dish data \citep{difrancesco2008}
presented in contours.

\begin{figure*}[htb]
  \includegraphics[width=0.99\textwidth]{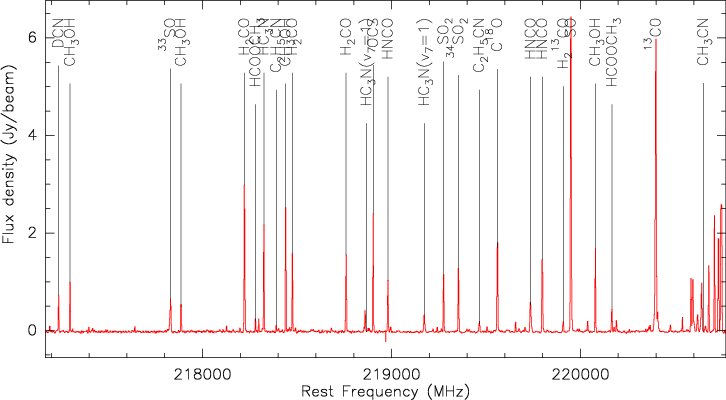}
  \caption{Example wide-band spectrum extracted toward AFGL2591. The
    most important lines in the bandpass are marked.}
\label{afgl2591_widex} 
\end{figure*} 

Regarding the evolutionary stage of the sample, they are all luminous
and massive young stellar objects (MYSOs) or otherwise named high-mass
protostellar objects (HMPOs).  Subdividing the regions a bit further,
some regions show very strong (sub)mm spectral line emission
indicative of hot molecular cores (AFGL2591, G75.78+0.34, CepAHW2,
W3(H$_2$O), NGC7538IRS1), other regions are line-poor (e.g., S87IRS1,
S106, G100.3779, G084.9505, G094.6028, G138.2957, G139.9091), and the
remaining sources exhibit intermediate-rich spectral line
data. Furthermore, the sample covers various combinations of
associated cm continuum, H$_2$O and class II CH$_3$OH maser emission
(Table \ref{sample}). Following \citet{motte2007}, we checked whether
the sources belong to the so-called IR-bright or IR-quiet categories
with the dividing line defined as IR-quiet when $S_{21\mu m} <
10\rm{Jy} \left(\frac{1.7\rm{kpc}}{D}\right)^2
\left(\frac{L}{1000\rm{L}_{\odot}}\right)$.  In contrast to our
initial expectation that all sources would classify as IR-bright, we
clearly find some diversity among the sample (see Table
\ref{sample}). While the majority indeed qualified as IR-bright, a few
sources fall in the IR-quiet category. Maybe slightly surprising, a
few of our line-brightest sources are categorized as IR-quiet (e.g.,
CepA and W3(H$_2$O)). Therefore, the differentiation in these two
categories only partly implies that the IR-quiet sources are
potentially younger, but it suggests at least that these sources are
still very deeply embedded into their natal cores. In this embedded
stage, they are already capable of driving dynamic outflows, have high
luminosities and produce a rich chemistry. 

A different evolutionary time indicator sometimes used is the
luminosity-over-mass ratio $L/M$ of the regions (see Table
\ref{sample}, e.g.,
\citealt{sridha,molinari2008,molinari2016,ma2013,cesaroni2017,motte2017}). The
CORE sample covers a relatively broad range in this parameter space
between roughly 20 and 700\,L$_{\odot}$/M$_{\odot}$. However, this
ratio is not entirely conclusive either. For example, the region with
our lowest ratio (S87IRS1 with $L/M\sim 18$\,L$_{\odot}$/M$_{\odot}$),
that could be indicative of relative youth, is classified otherwise as
IR-bright which seems counterintuitive at first sight. Since the
various age-indicators are derived from parameters averaged over
different scales, it is possible that they are averaging over
sub-regions with varying evolutionary stages and are hence not giving
an unambiguous evolutionary picture.

In summary, the CORE sample consists of regions containing HMPOs/MYSOs
above $10^4$\,L$_{\odot}$ from the pre-hot-core stage to typical
hot-cores and also a few more evolved regions that have likely already
started to disrupt their original gas core. The evolutionary stages are
comparable to the sample by \citet{palau2013,palau2015} with the
difference that they had a large fraction of sources below
$10^4$\,L$_{\odot}$ and even below $10^3$\,L$_{\odot}$ (only four
regions above $10^4$\,L$_{\odot}$).

\section{CORE large program strategy} 
\label{strategy}

Based on our experience with NGC7538IRS1 and NGC7538S
\citep{beuther2012c,beuther2013b}, we devised the CORE survey in
a similar fashion. The full sample is observed in the 1.3\,mm band,
and a sub-sample of five regions will also subsequently be observed at
843\,$\mu$m. Here we focus on the 1.3\,mm part of the
survey for the full sample. The shorter wavelength study will be
presented after its completion.

Several aspects were considered to achieve the goals of the project:
(i) The most extended A-configuration of NOEMA was used for the
highest possible spatial resolution (ii) Complementary observations
with more compact configurations of the interferometer recover
information on larger spatial scales. Simulations showed that adding the B
and D configurations provided the best compromise between spatial
information and observing time. (iii) To also cover very extended
spectral line emission, short spacing observations from the IRAM 30m
telescope were added. (iv) Spectrally, among other lines our survey
covers CH$_3$CN to trace high-density gas as might be found in
accretion disks and/or toroids (e.g., \citealt{cesaroni2007}) and
H$_2$CO which traces lower-density, larger-scale structures.  Both,
CH$_3$CN and H$_2$CO are also well known temperature tracers (e.g.,
\citealt{mangum1993,zhang1998b,araya2005}). Furthermore, outflow
tracers like $^{13}$CO and SO are included. A plethora of additional
lines are also covered to investigate the chemical properties of the
regions. An early example of such investigation can be found in the
paper about the pilot study sources NGC7538IRS1 and NGC7538S by
\citet{feng2016c}.

\begin{table}[htb]
\caption{Spectral lines at high spectral resolution}
\begin{tabular}{lrr}
\hline \hline
Line & $\nu$ &  $E_u/k$ \\
     & (GHz) &   (K) \\
\hline
H$_2$CO$(3_{0,3}-2_{0,1})$      &218.222 & 21 \\
HCOOCH$_3(17_{3,14}-16_{3,13})$  &218.298 & 100  \\
HC$_3$N$(24-23)$              & 218.325 & 131 \\
CH$_3$OH$(4_{2,2}-3_{1,2})$     &218.440 &  46 \\
NH$_2$CHO$(10_{1,9}-9_{1,8})$   &218.460 &  61 \\
H$_2$CO$(3_{2,2}-2_{2,1})$      &218.476 &  68 \\
OCS$(18-17)$                 & 218.903 & 100 \\
HCOOCH$_3(17_{4,13}-16_{4,12})$ &220.167 &  103 \\
CH$_2$CO$(11_{1,11}-10_{1,10})$ &220.178 &  77 \\
HCOOCH$_3(17_{4,13}-16_{4,12})$ &220.190 &  103  \\
CH$_3$CN$(12_6-11_6)$         &220.594 &  326 \\
CH$_3^{13}$CN$(12_3-11_3)$     &220.600 &  133 \\
CH$_3^{13}$CN$(12_2-11_2)$     &220.621 &  98  \\
CH$_3$CN$(12_5-11_5)$         &220.641 &  248 \\
CH$_3$CN$(12_4-11_4)$         &220.679 &  183 \\
CH$_3$CN$(12_3-11_3)$         &220.709 &  133 \\
CH$_3$CN$(12_2-11_2)$         &220.730 &  98 \\
CH$_3$CN$(12_1-11_1)$         &220.743 &  76 \\
CH$_3$CN$(12_0-11_0)$         &220.747 &  69 \\
\hline \hline
\end{tabular}
\label{linelist}
\end{table}

With the wide-band correlator units WIDEX, a spectral range from
217.167 to 220.834\,GHz was covered at a spectral resolution of
1.95\,MHz, corresponding to a velocity resolution of
$\sim$2.7\,km\,s$^{-1}$ at the given frequencies. Figure
\ref{afgl2591_widex} shows an example spectrum from AFGL2591. These
wide-band units are used to extract the line-free continuum as well as
to get a chemical census of the region. Furthermore, the velocity
resolution is sufficient for outflow investigations. However, to study
the kinematics of the central rotating structures, higher spectral
resolution is required. Therefore, we positioned the eight narrow band
correlator units to specific spectral locations covering the most
important lines at a spectral resolution of 0.312\,MHz, corresponding
to a velocity resolution of $\sim$0.43\,km\,s$^{-1}$ at the given
frequencies. Table \ref{linelist} shows the spectral lines covered at
this high spectral resolution. For more details about the spectral
line coverage we refer the reader to the CORE paper by Ahmadi et
al.~(subm.).

For the complementary IRAM 30\,m short spacings observations, we
mapped all regions with approximate map sizes of $1'$ in the
on-the-fly mode in the 1\,mm band. Since the bandpasses at the 30\,m
telescope are broader and the receivers work in a double-sideband
mode, the 30\,m data cover a broader range of frequencies between
$\sim$213 and $\sim$221\,GHz in the lower sideband and between
$\sim$229 and $\sim$236\,GHz in the upper sideband. The line data that
are covered by the NOEMA and 30\,m observations can be merged and
imaged together whereas the remaining 30\,m bandpass data can be used
as standalone data products. Since we do not use the single-dish data
for the continuum study presented here, we refer to the CORE paper by
Mottram et al.~(in prep.) for more details on the IRAM 30\,m data.

More details about the CORE project are provided at the team web-page
at http://www.mpia.de/core. There, we will also provide the final
calibrated visibility data and imaged maps. The data release will take
place in a staged fashion: the continuum data are published now, the
corresponding line data will be provided subsequently.

\section{Observations}
\label{obs}

\begin{figure}[htb]
  \includegraphics[width=0.49\textwidth]{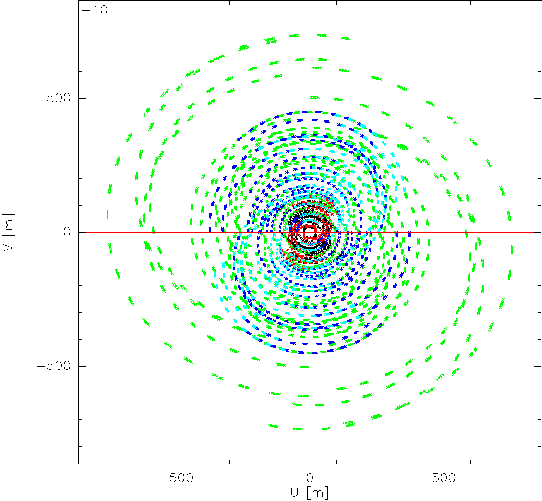}
  \caption{Example uv-coverage for CepA. The different colors
    correspond to different observed (half-)tracks. Red and black
    correspond to D-array observations, blue and cyan to B-array, and
    green to A-array data.}
\label{cepa_uv} 
\end{figure}

The entire CORE sample (except the pilot sources NGC7538IRS1 and
NGC7538S) was observed at 1.37\,mm between summer 2014 and January
2017 in the three PdBI/NOEMA configurations A, B and D to cover as
many spatial scales as possible (see section \ref{strategy}). The
baseline ranges for all tracks in terms of uv-radius are given in
Table \ref{para}. The shortest baselines, typically between 15 and
20\,m, correspond to theoretically largest recoverable scales of
$16''-20''$. For each track, two sources were observed together in a
track-sharing mode. The phase centers of each source and the
respective source pairs for the track-sharing are shown in Table
\ref{sample}. Since each source was observed in three different
configurations, at least three (half-) tracks were observed per
source. Depending on the conditions, several source pairs were
observed in more than three (partial) tracks in order to achieve the
required sensitivity and uv-coverage. Altogether, this
multi-configuration and multi-track approach resulted in excellent
uv-coverage for each source, an example of which is shown in
Fig.~\ref{cepa_uv}.  Typically two phase calibrators were observed in
the loops with the track-sharing pairs. For the final phase
calibration, we mostly only used the stronger ones. Depending on array
configuration and weather conditions, the phase noise varied between
$\sim$10 and $\sim50\deg$. Bandpass calibration was conducted with
observations of strong quasars, e.g., 3C84, 3C273, or 3C454.3. The
resulting spectral baselines are very good, over the broad WIDEX
bandpass as well as the narrow-band bandpasses (e.g., see
Fig.~\ref{afgl2591_widex}). The absolute flux calibration was
conducted in most cases with the source MWC349 where an absolute model
flux of 1.86\,Jy at 220\,GHz was assumed \footnote{MWC349 shows
  barely any variability at mm wavelength in continuous monitoring
  with NOEMA.}. For only very few tracks in which that source was not
observed, the flux calibration was conducted with other well-known
calibrators (e.g., LKH$\alpha$101). The absolute flux scale is
estimated to be correct to within 20\%.

\begin{table*}[htb]
\caption{CORE parameters}
\begin{center}
\begin{tabular}{lrrrrrrrrr|cc}
\hline 
\hline
Source & Beam & lin.~res.$^d$ & uv-radius$^e$ & rms & rms$_{\rm{sc}}$  & $5\sigma$ & $S_{\rm{peak}}$ & $S_{\rm{int}}$ & mf$^{a}$ & $T$(H$_2$CO) & $\Delta \varv$(H$_2$CO)\\
       & ($''$, PA) & (AU) & (m) & ($\frac{\rm{mJy}}{\rm{beam}}$) & ($\frac{\rm{mJy}}{\rm{beam}}$) & (M$_{\odot}$) & ($\frac{\rm{mJy}}{\rm{beam}}$)  & (mJy) & (\%) & (K) & (km\,s$^{-1}$)\\
\hline 
IRAS23151 & $0.45''\times 0.37''$(50$^{\circ}$) & 1350 & 21-764 & 0.19 & 0.10 & 0.05 & 32.6 & 100 & 78 & 59 & 3.4 \\
IRAS23033 & $0.45''\times 0.37''$(47$^{\circ}$) & 1760 & 20-765 & 0.46 & 0.28 & 0.28 & 38.9 & 310 & 64 & 55 & 3.5 \\
AFGL2591 & $0.47''\times 0.36''$(65$^{\circ}$)  & 1370 & 31-765 & 0.60 & 0.40 & 0.18 & 87.3 & 249 & 84 & 69 & 3.1 \\
G75.78 & $0.48''\times 0.37''$(60$^{\circ}$)    & 1615 & 21-765 & 0.60 & 0.42 & 0.16 & 64.7 & 256 & 87 & 108& 5.3 \\
S87IRS1 & $0.54''\times 0.35''$(37$^{\circ}$)   & 980  & 16-765 & 0.23 & 0.21 & 0.06 & 33.7 & 214 & 87 & 48 & 3.7 \\
S106 & $0.47''\times 0.34''$(47$^{\circ}$)      & 530  & 19-765 & 1.25 & 0.62 & 0.02 & 73.9 & 170 & 87 & 135& 4.8 \\
IRAS21078 & $0.48''\times 0.33''$(41$^{\circ}$) & 650  & 34-765 & 0.60 & 0.28 & 0.03 & 34.7 & 1020& 53 & 66 & 4.9 \\
G100 & $0.49''\times 0.33''$(56$^{\circ}$)      & 1440 & 16-765 & 0.08 & 0.05 & 0.03 & 8.5  & 67  & --$^b$ & 58 & 2.3 \\
G084 & $0.43''\times 0.38''$(69$^{\circ}$)      & 2230 & 15-753 & 0.10 & 0.08 & 0.22 & 6.2  & 85  & 67$^c$ & 35 & 3.5 \\
G094 & $0.41''\times 0.39''$(77$^{\circ}$)      & 1600 & 15-762 & 0.14 & 0.11 & 0.36 & 13.6 & 90  & 81 & 18 & 2.5 \\
CepA & $0.44''\times 0.38''$(80$^{\circ}$)      & 290  & 19-765 & 4.00 & 1.70 & 0.02 & 440.9& 1225& 72 & 119& 5.3 \\
NGC7538IRS9& $0.44''\times 0.38''$(80$^{\circ}$)& 1110 & 19-765 & 0.30 & 0.15 & 0.04 & 41.2 & 237 & 76 & 86 & 4.0\\
W3(H$_2$O) & $0.43''\times 0.32''$(86$^{\circ}$)& 750  & 19-760 & 4.50 & 1.90 & 0.13 & 451.6& 5292& 25 & 162& 6.6\\
W3IRS4 & $0.45''\times 0.32''$(83$^{\circ}$)    & 770  & 19-762 & 0.60 & 0.60 & 0.11 & 39.3 & 377 & 87 & 66 & 4.2\\
G108 & $0.50''\times 0.44''$(49$^{\circ}$)      & 2020 & 17-765 & 0.25 & 0.15 & 0.24 & 14.8 & 60  & --$^b$ & 36 & 3.3 \\
IRAS23385 & $0.48''\times 0.43''$(58$^{\circ}$) & 2230 & 18-764 & 0.25 & 0.11 & 0.11 & 18.0 & 190 & 56 & 73 & 3.8\\
G138 & $0.50''\times 0.41''$(60$^{\circ}$)      & 1320 & 20-764 & 0.16 & 0.16 & 0.12 & 6.2  & 100 & 82 & 36 & 2.9\\
G139 & $0.51''\times 0.40''$(56$^{\circ}$)      & 1460 & 21-764 & 0.17 & 0.15 & 0.10 & 13.9 & 26  & 95 & 48 & 1.4\\
\hline                                                                 
            & previous pilot study \\                                  
NGC7538IRS1&$0.33''\times 0.32''$(-55$^{\circ}$)& 880  & 68-765 & 10.0 & 5.20 & 1.34 & 2334 & 2838& 50 & 82 & 4.5 \\
NGC7538S & $0.34''\times 0.31''$(-81$^{\circ}$) & 880  & 68-765 & 0.60 & 0.50 & 0.14 & 28.1 & 253 & 91 & 78 & 5.6 \\
\hline
\hline
\end{tabular}
\end{center}
{\footnotesize ~\\
  The columns give the synthesized beam, the linear resolution, the baseline range (uv-radius), the rms noise before and after self-calibration, the $5\sigma$ mass sensitivity, the measured peak and integrated flux densities $S_{\rm{peak}}$ and $S_{\rm{int}}$, the missing flux ratios as well as the H$_2$CO derived temperatures $T$ and line widths $\Delta \varv$.\\
  $^a$ Missing flux, for details see main text\\
  $^b$ No single-dish data available\\
  $^c$ Based on BOLOCAM 1.1\,mm flux measurement in $40''$ aperture \citep{ginsburg2013}\\
  $^d$ Average linear resolution\\
  $^e$ Projected uv baseline range
}
\label{para}
\end{table*}

\begin{figure*}[htb]
  \includegraphics[width=0.99\textwidth]{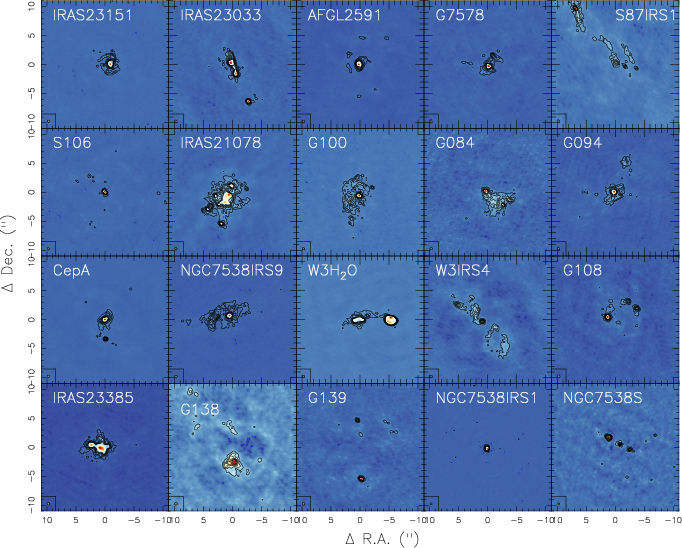}
  \caption{Compilation of 1.37\,mm continuum images for CORE sample on
    the same angular scale. The contouring is in $5\sigma$ steps (see
    Table \ref{para}). The sources are labeled in each panel, and the
    synthesized beams are shown at the bottom-left of each panel. A
    comparison figure converted to linear scales is shown in
    Fig.~\ref{contalllinear}. Zooms and absolute flux-scales are shown
    in Appendix \ref{individual_images}.}
\label{contall} 
\end{figure*} 

To achieve the highest angular resolution, uniform weighting was
applied during the imaging process. The final synthesized beams for
the continuum combining all NOEMA data vary between $\sim 0.32''$ and
$\sim 0.5''$ with exact values for each source given in Table
\ref{para}. The full width at half maximum of the primary beam of our
observations is $\sim$22$''$. To create the continuum images, we
carefully inspected the WIDEX bandpasses for each source individually
and created the continuum from the line-free parts only. The $1\sigma$
continuum rms correspondingly varies from source to source. This
depends not only on the chosen line-free channels, but also on the
side-lobe noise introduced by the strongest sources in the
fields. Although the uv-coverage is very good (Fig.~\ref{cepa_uv}),
not all side-lobes can be properly subtracted, and the final noise
depends on that as well.

To reduce calibration, side-lobe and imaging issues, we explored how
much self-calibration would improve the data quality. For that
purpose, we exported the continuum uv-tables to \textsc{casa} format
and did the self-calibration within \textsc{casa} (version 4.7.2,
\citealt{mcmullin2007}). We performed phase self-calibration only, and
the time intervals used for the process varied from source to source
depending on the source strength. Solution intervals of either 220,
100 or 45\,sec were used, where 45\,sec is the smallest possible
interval due to averaging of the data during data
recording. Interactive masking during the self-calibration loops was
applied, with only the strong peaks used in the first iterations and
then subsequently adapted to the weaker structures. After the
self-calibration, we again exported the data to \textsc{gildas} format
and conducted all the imaging within \textsc{gildas} to enable direct
comparisons with the original datasets. Again, uniform weighting was
applied and we cleaned the data down to a $2\sigma$ threshold. To show
the differences of the images prior to and after the self-calibration
process, Appendix \ref{individual_images} presents the derived images
before and after the self-calibration. The contouring is done in both
cases in $5\sigma$ steps. Careful inspection of all data shows that no
general structural changes were created during the self-calibration
process. The self-calibration improved the data considerably with
reduced rms noise and slightly increased peak fluxes. We find that the
flux-ratios between the main sub-structures within individual regions
remained relatively constant prior to and after self-calibration. In
the rest of the paper, we will conduct the analysis with the
self-calibrated dataset. Table \ref{para} presents the $1\sigma$
continuum rms for all sources before and after self-calibration. We
typically achieve sub-mJy rms with a range between 0.05 and
1.9\,mJy\,beam$^{-1}$ for the 18 new targets. Only the pilot source
NGC7538IRS1 has a slightly higher rms of $5.2\,$\,mJy\,beam$^{-1}$
which can be attributed to the higher source strength and the missing
D-array observations. Primary-beam correction was applied to the final
images, and the fluxes were extracted from these primary-beam
corrected data (section \ref{extraction}). Evaluating the measured
peak flux densities $S_{\rm peak}$ and noise values (rms$_{\rm sc}$)
in Table \ref{para} we find signal-to-noise ratios between 39 and 326
with the majority of region (13) exhibiting signal-to-noise ratios
greater than 100. We are providing in electronic form the original
pre-self-calibration images, the images after applying
self-calibration as well as the primary-beam corrected images.

\paragraph{Simulated observations:} To better understand how the
imaging affects our results, we simulated a typical observation. The
details of the simulations can be found in Appendix
\ref{simulations}. To summarize the method and results: We used real
single-dish dust continuum data from the large-scale SCUBA-2
850\,$\mu$m map of Orion by \citet{lane2016}, converted the flux to
1.37\,mm wavelength (assuming a $\nu^{3.5}$ frequency-relation),
rescaled the spatial resolution and flux density to a distance of
3\,kpc, and imaged different parts of Orion with the typical
uv-coverage and integration time from the CORE project. Similar to our
observations, the rms varied depending on whether a strong source (in
this case Orion-KL) was present in the observed field. While the point
source mass sensitivity is very good, between 0.01 and
0.1\,M$_{\odot}$ (depending on the rms), with our spatial resolution
typical Orion cores are extended structures, rather than point
sources, even at a distance of 3\,kpc. Hence, the dependence of the
rms noise on the strongest sources in the field strongly affects the
actual core mass sensitivity for extended structures as well. Taking
the two examples shown in Appendix \ref{simulations}, cores with
masses down to $\sim$1\,M$_{\odot}$ are detectable in fields without
very strong sources. If such a low-mass core were within the stronger
Orion-KL field, it would not be detectable anymore. Therefore, the
core mass sensitivities strongly depend on the strongest and most
massive sources within the respective observed fields. The dynamic
range limit of the simulations of Orion-KL is approximately 53.

\section{Continuum structure and fragmentation results}
\label{results}

\begin{figure*}[htb]
  \includegraphics[width=0.99\textwidth]{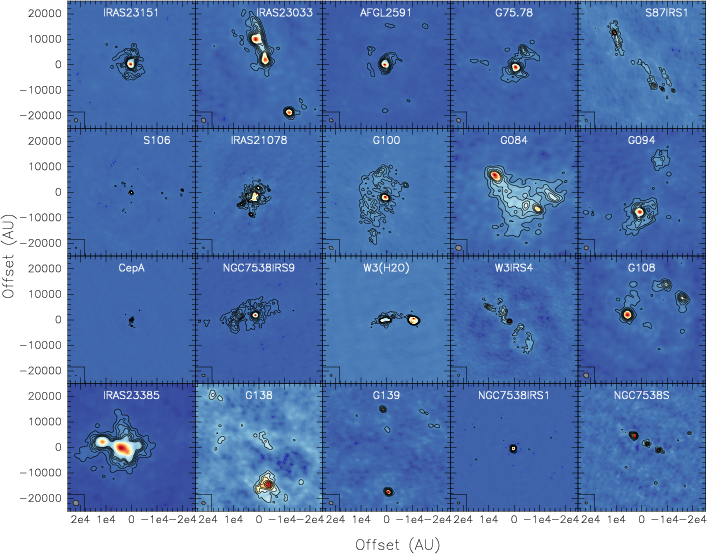}
  \caption{Compilation of 1.37\,mm continuum images for CORE sample
    converted to linear resolution elements. The contouring is in
    $5\sigma$ steps (see Table \ref{para}). The sources are labeled in
    each panel, and the synthesized beams are shown at the bottom-left
    of each panel. }
\label{contalllinear} 
\end{figure*} 

\subsection{Source structures}
\label{structure}

Figures \ref{contall} and \ref{contalllinear} present the 1.37\,mm
continuum data of the full CORE sample. While Fig.~\ref{contall} shows
the data in angular resolution over the full area of the primary beam
of the observations, Fig.~\ref{contalllinear} uses the distances of
the sources (Table \ref{sample}) and presents the data at the same
linear scales, making direct comparisons between sources possible. The
first impression one gets from these dust continuum images is that the
structures are far from uniform. While some sources are dominated by
single cores (e.g., IRAS23151, AFGL2591, S106, NGC7538IRS1), other
regions clearly contain multiple cores with a lot of substructures
(e.g., S87IRS1, IRAS\,21078, W3IRS4), some of which have more than 10
cores within a single observed field (see section
\ref{extraction}). We see no correlation between the number of
fragments and the distances to the sources. We will discuss this
fragmentation diversity in more detail in section \ref{fragmentation}.

\subsection{Source extraction}
\label{extraction}

To extract the sources from our 20 images, we used the classical {\sc
  clumpfind} algorithm by \citet{williams1994} on our self-calibrated
images. As input parameters we used the $5\sigma$ contour levels
presented in Figures \ref{contall} and \ref{contalllinear} as well as
in Appendix \ref{individual_images}. These images sometimes also show
negative $5\sigma$ contours, indicating that the interferometric noise
is neither uniform nor really Gaussian. Therefore, we inspected all
sources identified by {\sc clumpfind} individually and only included
those where the peak flux density is $\geq$10$\sigma$ (two positive
contours minimum in Appendix \ref{individual_images}).  The derived
positional offsets from the phase center, peak flux densities
$S_{\rm{peak}}$, integrated flux densities $S_{\rm{int}}$ and equivalent core
radii (calculated from the measured core area assuming a spherical
distribution) are presented in Table \ref{mass_flux} ($S_{\rm{peak}}$
and $S_{\rm{int}}$ are derived from the primary-beam corrected data).

To estimate the amount of missing flux filtered out by the
interferometric observations, we extracted the 850\,$\mu$m peak flux
densities from single-dish observations, mainly from the SCUBA legacy
archive catalogue \citep{difrancesco2008}. Since this dataset has a
final beam size of $22.9''$ it covers our primary beam size very
well. Scaling this 850\,$\mu$m data with a typical $\nu^{3.5}$
dependency to the approximate flux at our observing frequency of
220\,GHz, we can compare these values to the sum of the integrated
fluxes measured for each target region from our previous {\sc
  clumpfind} analysis. Table \ref{para} presents the corresponding
missing flux values (mf in percentage) for the sample (for two regions
-- G100 \& G108 -- we did not find corresponding single-dish
data). The amount of missing flux varies significantly over the
sample, typically ranging between 60 and 90\%. The only extreme
exception is W3(H$_2$O) where only 25\% of the flux is filtered
out. This implies that for this region the flux is strongly centrally
concentrated without much of a more extended envelope structure. For
the remaining sources, even with the comparably good uv-coverage
(Fig.~\ref{cepa_uv}) a significant fraction of the flux is filtered
out. The variations from source to source indicate that the spatial
density structure varies strongly from region to region as well (see
also discussion in section \ref{fragmentation}).

\begin{figure}[htb]
  \includegraphics[width=0.49\textwidth]{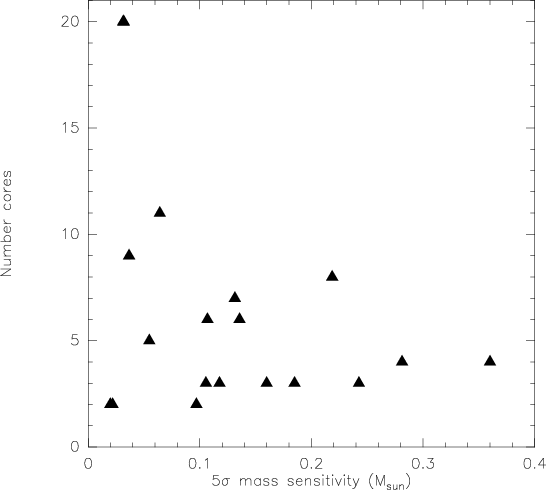}
  \caption{Number of identified cores plotted against the $5\sigma$
    mass sensitivity. NGC7538IRS1 is excluded because of its unusually
    high mass sensitivity limit).}
\label{masssens} 
\end{figure}

There is a broad distribution in the number of cores identified in
each region. We find between 1 and 20 cores among the different
regions (see Table \ref{min_span_tree}). To check whether this range
of identified cores is related to our mass sensitivity, in Figure
\ref{masssens} we plot the 5$\sigma$ mass sensitivity (Table
\ref{para} and section \ref{mass_calc}) versus the number of
identified cores (excluding NGC7538IRS1 because of its unusually poor
mass sensitivity limit, Table \ref{para}). While there might be a
slight trend of more cores towards lower mass sensitivity limits, our
lowest mass sensitivity limit region CepA also shows only two
cores. In the main regime of $5\sigma$ mass sensitivities between 0.1
and 0.3\,M$_{\odot}$ we do not see a relation between the number of
identified cores and the mass sensitivity. Hence, the number of
identified cores does not seem to be strongly dependent on our mass
sensitivity limits below 0.4\,M$_{\odot}$.

\subsection{Mass and column density distributions}
\label{mass_calc}

Assuming optically thin dust continuum emission at 220\,GHz, we can
estimate the gas masses and peak column densities for all identified
cores in the sample. Following the original outline by
\citet{hildebrand1983} in the form presented by \citet{schuller2009},
we use a gas-to-dust mass ratio of 150 \citep{draine2011}, a dust mass
absorption coefficient $\kappa$ of 0.9\,cm$^2$g$^{-1}$
(\citealt{ossenkopf1994} at densities of $10^6$\,cm$^{-3}$ with thin
ice mantles) and average temperatures for each region derived from the
CORE IRAM\,30m H$_2$CO data. H$_2$CO is a well-known gas thermometer
in the interstellar medium \citep{mangum1993}, and we derive
beam-averaged temperatures from the single-dish spectra toward the
peak positions of each region at a spatial resolution of $11''$. For
the temperature estimates we fitted the data with the \textsc{xclass}
tool (eXtended \textsc{casa} Line Analysis Software Suite) tool
\citep{moeller2017}. \textsc{xclass} models the spectra by solving the
radiative transfer equation for an isothermal homogeneous object in
local thermodynamic equilibrium (LTE), using the molecular databases
VAMDC and CDMS (http://www.vamdc.org and
\citealt{mueller2001}). \textsc{xclass} employs the model optimizer
package \textsc{magix} (Modeling and Analysis Generic Interface for
eXternal numerical codes) to find the best fit solutions
\citep{moeller2013}. The derived temperatures are shown in Table
\ref{para}.  Since we are deriving beam-averaged temperatures from the
single-dish data, the actual temperatures of individual cores at
smaller spatial scales may vary compared to that. More detailed
temperature analysis from the combined interferometer plus single-dish
data is beyond the scope of this paper and will be conducted in future
work on the CORE data. The mass estimates are in general lower limits
since we are filtering out large-scale flux that may be associated
with the dense cores (see also Appendix
\ref{simulations}). Furthermore, while the optically thin assumption
for the dust emission should be valid in most cases, there may be some
exceptions like CepA where high peak flux densities (Tables \ref{para}
\& \ref{mass_flux}) imply high brightness temperatures indicating
moderate optical depth at these peak positions. However, since the
masses are calculated typically over areas larger than just the peak,
and the brightness temperatures decrease quickly with distance from the
peak, this effect should be comparably weak.

The derived core masses and column densities are presented in Table
\ref{mass_flux} and roughly span 0.1 to 40\,M$_{\odot}$, and $5\times
10^{22}$ to $10^{25}$\,cm$^{-2}$. For the mass and column density
analysis, we excluded sources for which the continuum emission is
clearly dominated by H{\sc ii} regions and hence show barely dust
continuum emission. These are specifically W3(OH) (cores \#1 and \#2
in W3(H$_2$O), the southern ring-like region in W3IRS4 (sources \#5
and \#6) and core \#2 in S87IRS1. For several other cores, the fluxes
were corrected for free-free emission for the mass determinations (see
Table \ref{mass_flux}).

Using similar assumptions, we also re-estimated the large-scale mass
reservoir for the sample. For most sources, we used the integrated
850\,$\mu$m fluxes derived by \citet{difrancesco2008}, while for
IRAS\,23151 the 1.2\,mm flux was derived from the MAMBO data presented
in \citet{beuther2002a}, and for G084 we used the 1.1\,mm BOLOCAM data
from \citet{ginsburg2013}. The used gas-to-dust mass ratio and average
H$_2$CO derived temperatures are the same as above, and we used for
the single-dish data dust absorption coefficients $\kappa$ of 0.78,
0.9 and 1.4\,cm$^2$g$^{-1}$ at 1.2, 1.1 and 0.85\,mm wavelengths,
respectively (\citealt{ossenkopf1994} at densities of
$10^5$\,cm$^{-3}$). The derived total masses are presented in Table
\ref{sample} (for G100 and G108, the masses are taken from
C$^{18}$O(3--2) data from \citealt{maud2015b}). While the regions have
typical mass reservoirs of several 100\,M$_{\odot}$, the sample spans
a comparably broad range between $\sim$40 and $\sim$1500\,M$_{\odot}$
(for G108 even higher masses are measured, however over a comparably
large area with radius 1.4\,pc, Table \ref{sample} and
\citealt{maud2015b}).

\begin{figure}[htb]
  \includegraphics[width=0.49\textwidth]{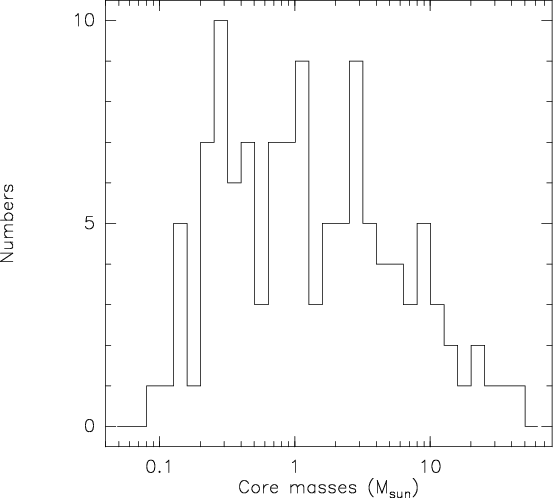}
  \includegraphics[width=0.49\textwidth]{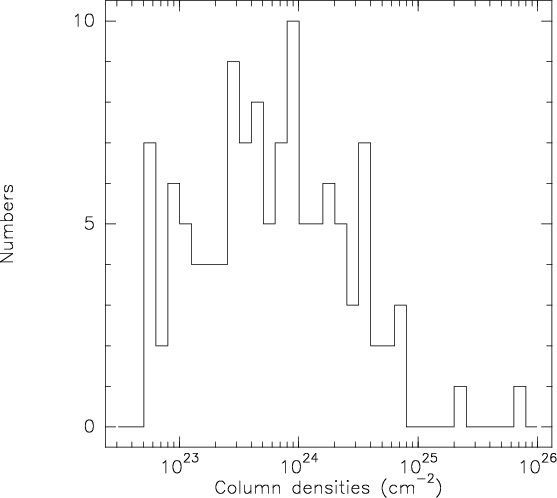}
  \caption{Histograms of masses (top panel) and column densities
    (bottom panel) for all detected cores.}
\label{histo} 
\end{figure} 

\begin{figure}[htb]
  \includegraphics[width=0.49\textwidth]{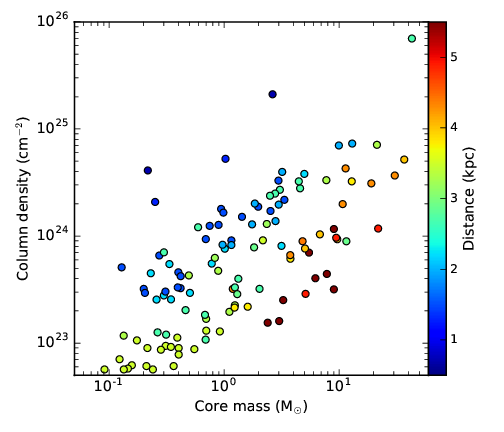}
  \caption{Gas column densities versus masses for all detected
    cores. The color-coding shows the distances of the sources.}
\label{mass_col} 
\end{figure}

For the NOEMA-only derived core parameters, Figure \ref{histo} shows
histograms of the masses and column densities. The combined mass
distribution shows that most detected cores are in the range between
$\sim$0.1 and $\sim$10\,M$_{\odot}$ with only a few cores exceeding
10\,M$_{\odot}$. The most massive core is in NGC7538IRS1 with
43\,M$_{\odot}$ (although significant free-free contamination may
affect the estimate for this source,
\citealt{beuther2012c}). Regarding the cores in excess of
10\,M$_{\odot}$, there is no clear trend whether they are found as
isolated objects or embedded in fragmented regions. For example,
comparably massive cores are found in the low-fragmentation regions
NGC7538IRS1 or AFGL2591, but cores of similar mass are also found in
more fragmented regions like IRAS\,23151, IRAS\,23033, G75.78, as well
as in the intermediately fragmented region W3(H$_2$O). The peak column
densities are very large, typically exceeding $10^{23}$\,cm$^{-2}$ and
even going above $10^{25}$\,cm$^{-2}$ for a few exceptional
regions. Figure \ref{mass_col} plots the column densities against the
masses, and while we see a scatter, there remains nevertheless a trend
that column densities and masses are correlated. If one takes into
account the distance-dependencies of our derived parameters
(color-coding in Fig.~\ref{mass_col}), we see that the
higher-mass-lower-column-density sources are found on average at
larger distances. With increasing distance the physical size of the
beam, where the column density is measured within, increases as
well. Such larger area beams cover the central highest-column-density
peak position but also more lower-column-density environmental
gas. This smoothing slightly decreases the measured column densities
with increasing distance. The other way round, increasing the covered
area with distance also increases the measured masses. Hence, part of
the scatter in Fig.~\ref{mass_col} is caused by the distance range of
our sample. For smaller distance bins, the scatter is significantly
reduced.

\begin{figure}[htb]
  \includegraphics[width=0.49\textwidth]{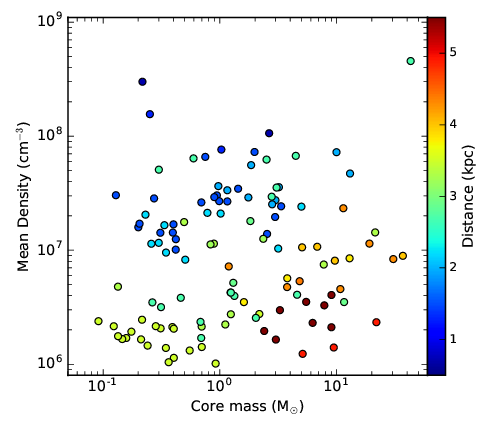}
  \caption{Mean core densities versus masses for all detected
    cores. The color-coding shows the distances of the sources.}
\label{mass_dens} 
\end{figure}

\begin{figure}[htb]
  \includegraphics[width=0.49\textwidth]{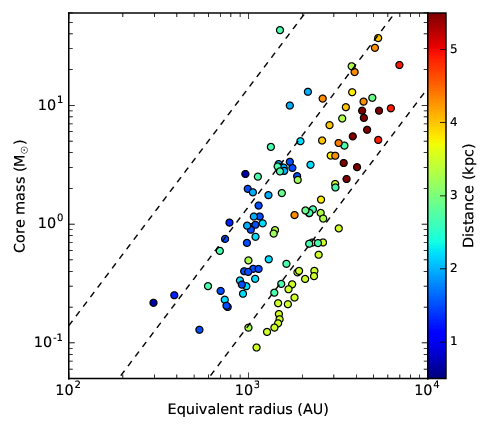}
  \caption{Core masses versus equivalent radii for all detected
    cores. The color-coding shows the distances of the sources. The
    dashed lines show constant column density with levels of
    $10^{23}$, $10^{24}$ and $10^{25}$\,cm$^{-2}$ from right to left.}
\label{mass_size} 
\end{figure}

Using the derived equivalent radii of the cores from the {\sc
  clumpfind} analysis (Table \ref{mass_flux}), we can also derive mean
densities for all cores under the assumption of spherical
symmetry. Figure \ref{mass_dens} plots these mean densities against
the corresponding core masses, again color-coded with distance. While
these average densities are rather high, typically between $10^6$ and
$10^8$\,cm$^{-3}$, there is no clear trend between the densities and
the masses. Taking again the distances into account the scatter is
reduced but identifying trends within distance-limited ranges is
still difficult. Hence, in this sample, the core densities are similar
over the whole range of observed core masses. Having a correlation
between mass and column density but less good correlation between mass
and average density implies that the core masses should correlate with
their sizes, i.e., equivalent radii. Figure \ref{mass_size} presents
the corresponding data again color-coded with distance. And indeed
mass and size are well correlated for the sample, again much tighter
if one looks at limited distance ranges. Figure \ref{mass_size} also
plots lines of constant column densities between $10^{23}$ and
$10^{25}$\,cm$^{-2}$. While most regions scatter between the $10^{23}$
and $10^{24}$\,cm$^{-2}$ lines, also sub-samples between limited
distance ranges do not follow constant column density distributions
but increase in column density with increasing mass, as already shown
in Fig.~\ref{mass_col}.

To estimate the typical Jeans fragmentation lengths and masses for the
clump scales, we assume mean densities of the original larger-scale
parental gas clumps between $10^5$ and $10^6$\,cm$^{-3}$ (e.g.,
\citealt{beuther2002a,palau2014}) and a temperature range between 20
and 50\,K, typical for regions in the given evolutionary stages. For
such conditions, the estimated Jeans length is between $\sim$5500 and
27700\,AU. For comparison, the corresponding Jeans masses in this
parameter range vary between 0.3 and 3.5\,M$_{\odot}$. While a large
fraction of the core masses lies within the regime of the Jeans
masses, a non-negligible number of sources also have higher masses
($\sim 36\%$) in excess of the Jeans mass of the original cloud. Since
our mass estimates are lower limits, even more cores may exceed the
estimated Jeans masses. However, since the mass estimates are affected
by many uncertainties (in addition to the missing flux, the assumed
dust properties and temperatures are adding an uncertainty of factors
2-4), the core separations may be a better proxy for analyzing the
fragmentation properties of the gas clumps.

\begin{figure}[htb]
  \includegraphics[width=0.49\textwidth]{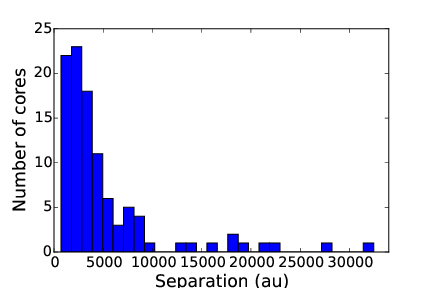}
  \caption{Nearest neighbor separation histogram from minimum spanning
    tree analysis}
\label{core_sep} 
\end{figure} 

\subsection{Core separations}
\label{core_sep_sec}

To quantify the core separations in all 20 sample regions, we employed
the minimum spanning tree algorithms available within the
\verb+astroML+ software package \citep{vanderplas2012} which
determines the shortest distances that can possibly connect each of
the cores in the sampled field. From this, the minimum, maximum and
mean separations of the cores in each field were determined, and are
presented in Table \ref{min_span_tree}, with the distribution of
nearest neighbor separations shown in Figure \ref{core_sep}. Since our
data are 2D projections of 3D distributions, these measured
separations are necessarily lower limits. The minimum core separations
are typically on the order of a few 1000\,AU (peak at $\sim$2000\,AU,
similar to \citealt{palau2013}) with only a few core separations for
the most nearby sources being measured below 1000\,AU. However, this
lower limit is most likely not a real physical lower separation limit
but associated with the spatial resolution. With typical resolution
elements around $0.3''-0.4''$ (Table \ref{para}) at distances of
several kpc (Table \ref{sample}), the linear spatial resolution is
below 1000\,AU for the most nearby sources (Table \ref{para}).

In contrast to likely not resolving all sub-structures within the
regions, we nevertheless observe strong fragmentation in many
targets. In particular, given the above estimated Jeans length between
$\sim$5500 and 27700\,AU (depending on density and temperature), most
regions appear to fragment at or below this thermal Jeans length
scale. Alternatively, the cores could have initially fragmented on
Jeans length scales, and then the fragments could have approached each
other even further due to the ongoing bulk motions from the global
collapse of the regions. In contrast to that, the turbulent Jeans
analysis, which includes the turbulent contributions to the sound
speed, results in significantly larger mass and length scales (e.g.,
\citealt{pillai2011,wang2014}) than the classical thermal Jeans
analysis. 

\begin{table}
\caption{Linear Minimum Spanning Tree Analysis}
\begin{tabular}{lccccc}
\hline \hline
Source & \#cores& mean sep & min sep & max sep \\
       &            & (AU)      & (AU)     & (AU) \\
\hline\hline
IRAS23151   & 5 & 3763   & 2195  & 5264  \\
IRAS23033   & 4 & 12185  & 5124  & 22616  \\
AFGL2591    & 3 & 15012  & 8284  & 21739  \\
G75.78      & 4 & 4392   & 3202  & 5924  \\
S87IRS1     & 11& 4564   & 1728  & 18625  \\
S106        & 2 & 5029   & 5029  & 5029  \\
IRAS21078   & 20& 1482   & 710   & 2491  \\
G100.3779   & 20& 3027   & 1573  & 7247  \\
G084.9505   & 8 & 6810   & 4247  & 9406  \\
G094.6028   & 4 & 9175   & 4521  & 18397  \\
CepAHW2     & 2 & 2382   & 2382  & 2382  \\
NGC7538IRS9 & 9 & 3087   & 1558  & 4524  \\
W3H2O       & 7 & 2583   & 1410  & 6071  \\
W3IRS4      & 6 & 3785   & 1069  & 7298  \\
G108.7575   & 3 & 13774  & 8341  & 19206  \\
IRAS23385   & 3 & 7413   & 6918  & 7909  \\
G138.2957   & 3 & 22088  & 16537 & 27640  \\
G139.9091   & 2 & 32468  & 32468 & 32468  \\
NGC7538IRS1 & 1 & \\
NGC7538S    & 6 & 7828   & 1520  & 13663  \\
\hline
\hline
\end{tabular}
\label{min_span_tree}
\end{table}

\section{Discussion}
\label{fragmentation}

Fragmentation occurs in general on various spatial scales and is
likely a hierarchical process. Within our CORE project, we investigate
the fragmentation processes on clump scales in high-mass star-forming
regions. We concentrate on the dense central structures on scales
above $\sim$1000\,AU and roughly below 50000\,AU or 0.25\,pc. These
largest scales correspond roughly to the largest theoretically
recoverable scales with 15\,m baselines at 3\,kpc distance (section
\ref{obs}). In the continuum study presented here we investigate the
fragmentation of clumps into cores. Fragmentation on smaller disk-like
scales will also be investigated by the CORE program, however, that is
more strongly based on the spectral line data and will be discussed in
complementary papers (e.g., Ahmadi et al.~subm., Bosco et al.~in
prep.).

\subsection{Thermal versus turbulent fragmentation}

With respect to the fragmentation of massive gas clumps, some
important questions are: What controls the fragmentation properties of
high-mass star-forming clumps? Is thermal Jeans fragmentation
sufficient? How important are additional parameters like an initial
non-uniform density profile or the magnetic field properties? How
important is global accretion onto the clump from the diffuse ISM?

Regarding turbulent and thermal contributions, a number of studies
have investigated this problem.  For example, \citet{wang2014} found
that the observed masses of fragments within massive infrared dark
cloud clumps are often more than 10\,M$_{\odot}$. These masses are an
order of magnitude larger than the thermal Jeans mass of the
clump. Therefore they argue that the massive cores in a protocluster
are more consistent with turbulent Jeans fragmentation (i.e.,
including a turbulent contribution to the velocity
dispersion). Similar results were found by \citet{pillai2011} in their
study of two young pre-protocluster regions. On the other hand,
\citet{palau2013,palau2014,palau2015} found in their compiled sample
of more evolved (IR-bright) star-forming regions that the masses of
most of the fragments are comparable to the expected thermal Jeans
mass, while the most massive fragments have masses a factor of 10
larger than the Jeans mass. \citet{palau2013,palau2014,palau2015}
concluded that these objects are consistent with thermal Jeans
fragmentation of the parental cloud, in agreement with recent other
investigations (e.g., \citealt{henshaw2017,cyganowski2017}). Recent
ALMA studies of regions containing hypercompact H{\sc ii} regions also
show small fragment separation scales \citep{klaassen2018}. In
addition to this, \citet{fontani2016} argue that the magnetic field is
important for the fragmentation of IRAS\,16061–5048c1 (see also
\citealt{commercon2011,peters2011}).

\begin{figure}[htb]
  \includegraphics[width=0.49\textwidth]{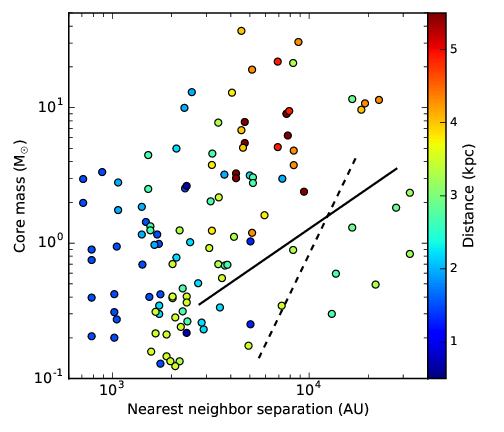}
  \caption{Fragment masses against nearest neighbor separation from
    the minimum spanning tree analysis. The full line corresponds to
    the Jeans lengths and Jeans masses calculated at 50\,K for a
    density grid between $10^5$ and $10^7$\,cm$^{-3}$. For comparison,
    the dashed line corresponds to the Jeans lengths and Jeans masses
    calculated at a fixed density of $5\times 10^5$\,cm$^{-3}$
    \citep{beuther2002a} with temperatures between 10 and 100\,K. The
    color-coding shows the distances of the sources.}
\label{sep_mass} 
\end{figure}

In our sample of high-mass star-forming regions, including regions in
an evolutionary stage comparable to those studied by
\citet{palau2013,palau2014,palau2015}, we find that most of the
fragment masses approximately agree with a plausible range of Jeans
masses, and most nearest-neighbor separations are below the predicted
scales of thermal Jeans fragmentation. To explore that in more detail,
Fig.~\ref{sep_mass} plots the derived core masses against the nearest
neighbor separation derived from the minimum spanning tree
analysis. The full and dashed lines show the relation between both
thermal Jeans mass and Jeans length depending on density and
temperature. In general, we do not find a clear trend between the two
properties, and distance does not seem to be the primary factor in the
observed scatter either. The Figure also shows that for the plausible
range of densities and temperatures ($10^5$ to $10^7$\,cm$^{-3}$ and
10 to 100\,K) the observed parameters are difficult to explain. One
has to keep in mind that both observables are lower limits: the mass
because of missing flux and the separation because of projection
effects. Accounting for these effects, the measurements could shift a
bit closer to the predicted lines, but could also shift sources
parallel to them. For comparison, in the turbulent Jeans fragmentation
picture, the sound speed is replaced by the velocity dispersion (e.g.,
\citealt{wang2014}), which is typically a factor 5 to 10 higher than
the thermal sound speed (see H$_2$CO line width $\Delta v$(H$_2$CO) in
Table \ref{para}). Even if not all the observed line width is caused
by pure turbulent motions, but also has contributions from organized
motions due to, e.g., large-scale infall, the regions clearly exhibit
turbulent motions. Since the Jeans length and mass depend to the first
and third power on the sound speed, respectively, replacing the
thermal sound speed with the turbulent sound speed would shift the
drawn correlations in Fig.~\ref{sep_mass} largely outside the observed
box beyond the top-right corner. While we cannot conclude that thermal
fragmentation explains everything, our data seem to refute that a
turbulent contribution is needed if one applies a simple Jeans
analysis for these spatial scales.

Several factors contributed to the apparent difference in
fragmentation analysis between \citet{wang2014} or \citet{pillai2011}
on the one side, and \citet{palau2013,palau2014,palau2015} and the
study here on the other side. First, the \citet{wang2014} sample,
incorporating data from \citet{zhang2009} and \citet{zhang2011}, has a
typical $1 \sigma$ mass sensitivity of 1\,$M_\odot$. Therefore, lower
mass fragments close to the global Jeans mass were not detected in
these observations. Indeed, more sensitive observations from ALMA
toward one of the objects in the sample, IRDC G28.34, revealed lower
mass fragments \citep{zhang2015}. Secondly, time evolution must play a
role since fragmentation is a continuous process. As mentioned in
section \ref{core_sep_sec}, the separation scales between fragments
may also change with evolutionary time. In the picture of globally
collapsing clouds and gas clumps, one would expect larger fragment
separation at early evolutionary stages. Then, during the ongoing
collapse, the fragments may move closer together, following the overall
gravitational contraction of the region. Therefore, the observed state
of fragmentation only represents a snapshot in the time evolution.  The
less evolved regions such as those in \citet{wang2014} or
\citet{pillai2011} may present a deficit of low-mass fragments because
the typical density of the cloud/clump is still lower so that a
distributed low-mass protostar population may not have formed yet
(e.g., \citealt{zhang2015}). Furthermore, the more evolved objects
such as those in this paper here have higher densities
(Fig.~\ref{mass_dens}), and therefore experience more fragmentation
and are potentially more advanced in forming low-mass protostars.

In addition to the presented fragmentation properties, we point out
that the nearest separations of cores are peaking around the spatial
resolution limit of the observation (Fig.~\ref{core_sep}). Hence,
fragmentation is also expected on even smaller scales. This can be
investigated for this sample by higher spatial resolution observations
with the future upgraded NOEMA (the baselines lengths are expected to
be doubled), and for more southern sources with ALMA.

Recently, \citet{csengeri2017} reported limited fragmentation for
earlier evolutionary stages based on Atacama Compact Array data at
$3.5''-4.6''$. At the given spatial resolution and a mass sensitivity
$>11$\,M$_{\odot}$ they find that in 77\% of their sample only three
or fewer massive cores are found. However, because of the lower
angular resolution and worse mass sensitivity, a direct comparison
between their and this study is not possible. The data of
\citet{csengeri2017} are complemented with ALMA 12\,m array data, and
the combined dataset will be very valuable for comparison with the
CORE project.

A different aspect to be considered is that the fragmentation
properties likely change with spatial
scale. \citet{kainulainen2013,kainulainen2017} have shown for two
filaments (the infrared dark cloud G11.11 and the Orion integral shape
filament) that the fragmentation properties appear to show distinct
signatures at different spatial scales. In particular for the infrared
dark cloud, \citet{kainulainen2013} argue that filament fragmentation
dominates on large spatial scales ($\geq 1$\,pc), whereas on smaller
spatial scales thermal Jeans fragmentation takes over ($\sim
0.2$\,pc). With respect to our CORE sample, analyzing the filamentary
properties on larger spatial scales is beyond the scope of this paper.
However, it is clear that the CORE study deals with massive
star-forming regions at high densities and not with the larger-scale,
potentially filamentary clouds. In relation to the work by
\citet{kainulainen2013,kainulainen2017}, we are in the second regime
that would be dominated by Jeans fragmentation. Therefore, our general
result that the CORE sample is more consistent with thermal Jeans
fragmentation is in agreement with the results by
\citet{kainulainen2013,kainulainen2017}.

\subsection{Fragmentation diversity} 

Our sample clearly shows that the fragmentation properties within
high-mass star-forming regions are not uniform, finding a diversity
from highly fragmented regions to those that host one or only very few
cores (see also \citealt{bontemps2010,palau2013,csengeri2017}). While
this sample seems in general largely consistent with thermal Jeans
fragmentation (see previous subsection), it should be noted that we
also find a few massive cores in excess of 10\,M$_{\odot}$
(sec.~\ref{mass_calc} and Fig.~\ref{histo}). A high level of
fragmentation with many low-mass cores favors high-mass star formation
scenarios in the framework of competitive accretion (e.g.,
\citealt{clark2006,bonnell2007,smith2009a}), whereas individual
massive cores are more strongly needed in the turbulent core picture
(e.g., \citealt{mckee2003,tan2014}). Because we find examples for both
pictures in our CORE sample, this may indicate that different
high-mass star formation scenarios are possible or even interplay with
each other.

Since the sample is selected to host high-mass protostellar objects
(HMPOs), the range of evolutionary stages is not broad. Nevertheless,
as discussed in section \ref{sec_sample}, within the HMPO category, we
cover regions with varying IR-brightness and luminosity-to-mass
ratios. Hence, while evolution is unlikely to be the main explanation
for the observed fragmentation diversity, it cannot be entirely
excluded. Furthermore, as discussed in the previous subsection,
different levels of initial turbulence are also unlikely to be the
underlying cause. Other possibilities to explain the different levels
of fragmentation are variations in the initial density profiles and/or
variations in the magnetic field properties. Differences in the
density profiles could also arise from environmental effects like
global collapse where the central gas clumps are continuously fed by
some larger-scale cloud envelope.

Since the whole sample is observed with rather uniform uv-coverages,
one wonders whether the amount of missing flux may be related to
density structure of the parental gas clump and by that to the
observed fragmentation properties of the cores. Therefore, we compare
a few extreme cases: The two comparably isolated regions AFGL2591 and
NGC7538IRS1 (both at similar distances at 3.3 and 2.7\,kpc, Table
\ref{sample}) show very different amounts of missing flux with values
of 84\% and 50\% of the flux being filtered out. At the other extreme,
two highly fragmented sources like S87IRS1 and IRAS21078 (at distances
of 2.2 and 1.5\,kpc, Table \ref{sample}) also exhibit very different
values of 87\% and 53\% of flux being filtered out. Hence, the overall
fraction of flux being lost because of the interferometric
observations -- or rather the amount of mass in a diffuse,
larger-scale reservoir -- appears not to be an important issue for the
observed fragmentation differences.

\citet{girichidis2011} have shown with simulations of star-forming
regions how the density profile affects the level of fragmentation:
While flat profiles ($\rho \propto$constant) resulted in many
fragments, they find that density profiles like $\rho \propto r^{-2}$
(over cloud radii of $\sim$0.1\,pc) quickly lead to the formation of a
single object at the center where further fragmentation is
prohibited. In their simulations, the intermediate case with $\rho
\propto r^{-1.5}$ is also dominated by a central object but additional
fragments can form depending on their initial turbulence
field. Observations of the density profiles of high-mass gas clumps by
different groups typically find density slopes $\rho \propto
r^{-\alpha}$ with $\alpha$ between 1.5 and 2.6 (e.g.,
\citealt{beuther2002a,mueller2002,fontani2002,hatchell2003}). Furthermore,
\citet{palau2014} find a weak inverse trend between level of
fragmentation and steepness of density profile, i.e., less
fragmentation for steep density profiles. Hence, it seems reasonable
that a range of initial density profiles can at least partly explain
the observed diversity of fragmentation properties in our CORE
sample. In future work, we are going to follow up on that and will
investigate the density structure of the regions based on the
combination of single-dish data with the interferometer data in more
depth.

In addition to this, different magnetic field properties in the
parental gas clumps can cause similar effects. Typically, the ratio
between gravity and magnetic field is phrased in terms of the critical
mass-to-flux ratio (e.g., \citealt{tilley2007}). \citet{commercon2011}
modeled the collapse of high-mass star-forming regions with a range of
magnetic field strengths. While their low-magnetic field case results
in a larger number of fragments, the high-magnetic field case is
dominated by a central massive object (see also
\citealt{fontani2016}). Similarly, \citet{peters2011} also find
reduced fragmentation with increasing magnetic field strength. To
really differentiate whether the initial density profile and/or
magnetic field properties are the dominant reason explaining the
observed fragmentation diversity, we need to know the magnetic field
strength as well as the initial density profile. For two regions
within the CORE sample (W3(H$_2$O) and NGC7538IRS1) magnetic field
studies have already been conducted with the Submillimeter Array on
arcsecond resolution scale \citep{chen2012,frau2014}. The derived
magnetic field strengths are comparably high in both regions with 17.0
and 2.5\,mG, respectively. Since both regions exhibit very few or even
only one fragment, the observed high magnetic field values are
consistent with the low degree of fragmentation in these two
regions. Future investigations in this direction are anticipated for
the whole sample, which in particular will reveal whether regions with
a high degree of fragmentation have a lower magnetic field strength.

\section{Conclusions and Summary}
\label{conclusion}

With the goals of studying the fragmentation, disk formation, outflows
and chemical properties during the birth of the most massive stars, we
have conducted the IRAM NOEMA large program CORE, observing a sample
of 20 high-mass star-forming regions at $0.3''-0.4''$ resolution in
the 1.37\,mm continuum and spectral line emission. In this paper, we
present the survey scope, its main observational characteristics, the
sample selection and the overall goals of the project. More details
about the project as well as the first data release of the continuum
data are provided at http://www.mpia.de/core. For a first scientific
analysis of the data, we concentrated on the 1.37\,mm dust continuum
emission to investigate the fragmentation properties during early
high-mass cluster formation.

We observe diverse fragmentation morphologies ranging from regions
that are dominated by single high-mass cores to those that fragment
into up to 20 cores. Since the sample contains mainly high-mass
protostellar objects (although with some range of evolution within
that category), larger-scale evolutionary effects are unlikely to
explain all the differences. Observational artifacts like interferometric
missing flux or different physical resolution can also be ruled
out. The typical nearest neighbor separations peak below the thermal
Jeans length determined from estimates of the initial average cloud
density, indicating that thermal gravitational fragmentation is
sufficient to explain the main observed core separations, and that
additional turbulent contributions to the Jeans analysis are not
needed for this sample. The diversity between regions with few or only
one fragment versus those with many fragments may be explained by
differences in the initial density structures of the maternal gas
clumps (potentially caused by environmental effects like global gas
infall from a surrounding envelope) and/or variations in the initial
magnetic field configurations. Since the nearest neighbor separation
peaks around our spatial resolution limit, it is likely that further
fragmentation takes place on even smaller spatial scales. With NOEMA,
we will be able to address such questions for this northern hemisphere
sample in a few years when the available baseline lengths will be
doubled. Furthermore, ALMA observations of complementary southern
hemisphere sources will investigate these questions in even greater
depth.

Other scientific questions related to the disk formation, outflow
properties and chemical processes during the formation of high-mass
stars will be addressed by complementary CORE papers focusing on the
spectral line data.

\begin{acknowledgements} 
  This work is based on observations carried out under project number
  L14AB with the IRAM NOEMA Interferometer and the IRAM 30\,m
  telescope. IRAM is supported by INSU/CNRS (France), MPG (Germany)
  and IGN (Spain). This paper made use of information from the Red MSX
  Source survey database at
  http://rms.leeds.ac.uk/cgi-bin/public/RMS\_DATABASE.cgi which was
  constructed with support from the Science and Technology Facilities
  Council of the UK. HB, AA, JCM and FB acknowledge support from the
  European Research Council under the Horizon 2020 Framework Program
  via the ERC Consolidator Grant CSF-648505. RK acknowledges financial
  support via the Emmy Noether Research Programme funded by the German
  Research Foundation (DFG) under grant no. KU 2849/3-1. RGM
  acknowledges support from UNAM-PAPIIT program
  IA102817. TCs acknowledges support from the Deut\-sche
  For\-schungs\-ge\-mein\-schaft (DFG) via the SPP (priority
  programme) 1573 'Physics of the ISM'. ASM acknowledges support from
  Deutsche Forschungsgemeinschaft through grant SFB956 (subproject
  A6). AP acknowledges financial support from UNAM and CONACyT, M\'exico.

\end{acknowledgements}


\newpage
\longtab{5}{
\begin{longtable}{lccccccc}
\caption{\label{mass_flux} Continuum source parameters}\\
\hline\hline
\# & $\Delta x$ & $\Delta y$ & $S_{\rm{peak}}$ & $S_{\rm{int}}$ & $r$ & $M$ & $N$ \\
   & ($''$) & ($''$) & ($\frac{\rm{mJy}}{\rm{beam}}$) & (mJy) & (AU) & (M$_{\odot}$) & ($10^{24}$\,cm$^{-2}$) \\ 
\hline
\endfirsthead
\caption{continued.}\\
\hline\hline
\# & $\Delta x$ & $\Delta y$ & $S_{\rm{peak}}$ & $S_{\rm{int}}$ & $r$ & $M$ & $N$ \\
   & ($''$) & ($''$) & ($\frac{\rm{mJy}}{\rm{beam}}$) & (mJy) & (AU) & (M$_{\odot}$) & ($10^{24}$\,cm$^{-2}$) \\ 
\hline
\endhead
\hline
\endfoot
\multicolumn{8}{l}{IRAS23151}\\
  1  &  -0.89  &   0.07  &   32.6  &   70.8  &  3339  &    7.8  &   3.32  \\
  2  &  -1.26  &   1.63  &    2.2  &   11.3  &  2534  &    1.2  &   0.23  \\
  3  &  -0.44  &  -1.18  &    1.9  &   10.2  &  2617  &    1.1  &   0.20  \\
  4  &   0.15  &  -0.07  &    1.6  &    6.4  &  2273  &    0.7  &   0.17  \\
  5  &  -1.26  &   2.29  &    1.1  &    1.2  &  1002  &    0.1  &   0.12  \\
\hline
\multicolumn{8}{l}{IRAS23033}\\
  1  &   0.30  &   0.30  &   33.4  &  151.9  &  5081  &   30.5  &   3.68  \\
  2  &  -2.73  &  -6.36  &   38.9  &   56.8  &  2599  &   11.4  &   4.28  \\
  3  &  -0.59  &  -1.55  &   28.2  &   95.0  &  3915  &   19.1  &   3.10  \\
  4  &   0.44  &  -2.14  &    2.9  &    5.9  &  1811  &    1.2  &   0.32  \\
\hline
\multicolumn{8}{l}{AFGL2591}\\
  1  &   0.22  &   0.07  &   87.3  &  234.3  &  3770  &   21.4  &   7.11  \\
  2  &   1.92  &   1.92  &    5.6  &    9.6  &  1410  &    0.9  &   0.47  \\
  3  &  -4.58  &   4.58  &    5.1  &    5.3  &  1002  &    0.5  &   0.43  \\
\hline
\multicolumn{8}{l}{G7578}\\
  1  &  -0.15  &  -0.22  &   64.7  &  169.4  &  3794  &   12.9  &   3.24  \\
  2  &  -0.96  &   1.48  &   12.2  &   49.6  &  2881  &    3.8  &   0.61  \\
  3  &   0.67  &  -1.55  &    4.4  &   21.1  &  2544  &    1.6  &   0.22  \\
  4  &  -0.74  &   0.67  &    4.3  &   16.2  &  2185  &    1.2  &   0.21  \\
\hline
\multicolumn{8}{l}{S87IRS1}\\
  1  &   6.87  &   9.75  &   33.7  &   81.6  &  1953  &    5.0  &   3.81  \\
  2  &  -0.15  &   0.37  &    5.8  &   18.7  &  1584  &    0.0  &   0.00  \\
  3  &  -1.70  &  -0.30  &    4.8  &    5.5  &   899  &    0.3  &   0.55  \\
  4  &   6.13  &   7.61  &    7.2  &   51.7  &  2222  &    3.2  &   0.81  \\
  5  &  -3.62  &  -0.59  &    4.0  &    3.8  &   740  &    0.2  &   0.45  \\
  6  &   6.65  &  11.60  &    6.8  &   16.6  &  1203  &    1.0  &   0.76  \\
  7  &   0.22  &   1.55  &    2.6  &    8.3  &  1301  &    0.5  &   0.29  \\
  8  &  -2.51  &   1.77  &    2.5  &    4.9  &   975  &    0.3  &   0.28  \\
  9  &   7.24  &  10.64  &    4.9  &   12.8  &  1097  &    0.8  &   0.55  \\
 10  &  -2.00  &   2.36  &    2.3  &    5.7  &  1093  &    0.3  &   0.26  \\
 11  &  -3.03  &   0.59  &    2.3  &    4.2  &   935  &    0.3  &   0.25  \\
\hline
\multicolumn{8}{l}{S106}\\
  1  &   0.00  &   0.07  &  136.0  &  162.5  &   786  &    1.0  &   5.27  \\
  2  &   2.66  &   2.88  &    7.0  &    7.5  &   387  &    0.3  &   2.08  \\
\hline
\multicolumn{8}{l}{IRAS21078}\\
  1  &   0.67  &  -0.44  &   34.7  &  148.3  &  1761  &    3.0  &   3.29  \\
  2  &   1.40  &  -1.77  &   23.0  &  166.9  &  1705  &    3.3  &   2.18  \\
  3  &   1.03  &  -0.74  &   19.8  &   98.8  &   993  &    2.0  &   1.88  \\
  4  &   0.07  &   1.11  &   18.1  &  126.5  &  1872  &    2.5  &   1.71  \\
  5  &   4.36  &  -2.88  &   18.9  &   47.1  &  1039  &    0.9  &   1.79  \\
  6  &   2.88  &  -0.59  &   15.9  &   71.5  &  1142  &    1.4  &   1.51  \\
  7  &   1.92  &  -5.39  &   17.4  &   49.3  &  1096  &    1.0  &   1.65  \\
  8  &   2.36  &  -2.14  &   13.4  &   44.7  &  1033  &    0.9  &   1.27  \\
  9  &   2.00  &  -1.77  &   13.1  &   37.4  &   743  &    0.8  &   1.24  \\
 10  &   1.63  &   0.44  &    9.6  &   57.8  &  1158  &    1.2  &   0.91  \\
 11  &   3.62  &  -2.29  &    9.9  &   34.6  &   983  &    0.7  &   0.94  \\
 12  &   5.03  &  -3.10  &    6.9  &   13.6  &   702  &    0.3  &   0.66  \\
 13  &  -2.44  &   0.59  &    5.4  &    6.4  &   534  &    0.1  &   0.51  \\
 14  &   3.77  &  -1.11  &    4.8  &   20.0  &   949  &    0.4  &   0.46  \\
 15  &   2.66  &   0.89  &    4.4  &   20.9  &  1065  &    0.4  &   0.42  \\
 16  &  -1.48  &   1.26  &    3.4  &   20.9  &  1142  &    0.4  &   0.33  \\
 17  &   1.70  &  -3.40  &    3.5  &   19.8  &   999  &    0.4  &   0.33  \\
 18  &   3.10  &   0.37  &    3.4  &   10.0  &   768  &    0.2  &   0.32  \\
 19  &   1.63  &  -4.29  &    3.2  &   15.4  &   921  &    0.3  &   0.30  \\
 20  &   2.07  &  -3.77  &    3.1  &   10.2  &   756  &    0.2  &   0.29  \\
\hline
\multicolumn{8}{l}{G100}\\
  1  &   0.15  &  -0.59  &    8.5  &   17.3  &  3049  &    2.2  &   0.91  \\
  2  &   0.67  &   2.29  &    1.2  &    5.6  &  2610  &    0.7  &   0.13  \\
  3  &   1.63  &  -2.14  &    1.2  &    7.3  &  3194  &    0.9  &   0.13  \\
  4  &   1.11  &   1.92  &    1.1  &    3.1  &  1880  &    0.4  &   0.11  \\
  5  &  -2.14  &   0.81  &    1.0  &    1.4  &  1482  &    0.2  &   0.11  \\
  6  &   2.36  &  -1.11  &    0.9  &    2.5  &  1756  &    0.3  &   0.09  \\
  7  &   1.63  &  -1.26  &    0.8  &    3.2  &  1919  &    0.4  &   0.09  \\
  8  &   2.00  &  -0.81  &    0.8  &    1.7  &  1467  &    0.2  &   0.09  \\
  9  &   1.11  &   0.81  &    0.8  &    4.4  &  2468  &    0.6  &   0.09  \\
 10  &   1.48  &  -4.36  &    0.9  &    2.7  &  2074  &    0.3  &   0.09  \\
 11  &   2.59  &  -2.07  &    0.8  &    2.3  &  1676  &    0.3  &   0.09  \\
 12  &  -0.30  &   0.30  &    0.7  &    3.2  &  2336  &    0.4  &   0.08  \\
 13  &   2.66  &  -2.66  &    0.7  &    1.0  &  1273  &    0.1  &   0.07  \\
 14  &  -0.74  &   0.81  &    0.6  &    2.9  &  2325  &    0.4  &   0.06  \\
 15  &   2.07  &   1.55  &    0.6  &    1.3  &  1495  &    0.2  &   0.06  \\
 16  &   2.59  &   0.30  &    0.6  &    1.7  &  1663  &    0.2  &   0.06  \\
 17  &   2.07  &   0.44  &    0.5  &    1.2  &  1467  &    0.1  &   0.06  \\
 18  &   1.48  &  -0.44  &    0.5  &    1.9  &  1811  &    0.2  &   0.06  \\
 19  &   2.29  &   0.96  &    0.5  &    1.1  &  1399  &    0.1  &   0.06  \\
 20  &   2.51  &   1.48  &    0.5  &    0.7  &  1112  &    0.1  &   0.06  \\
\hline
\multicolumn{8}{l}{G084}\\
  1  &   0.38  &   0.23  &    6.2  &   16.5  &  4311  &    9.0  &   1.16  \\
  2  &  -2.70  &  -2.18  &    3.8  &   10.0  &  3824  &    5.5  &   0.70  \\
  3  &  -1.88  &  -1.95  &    2.4  &   14.3  &  4410  &    7.8  &   0.44  \\
  4  &  -3.90  &  -1.43  &    2.2  &   11.4  &  4595  &    6.2  &   0.40  \\
  5  &  -0.45  &  -0.90  &    1.7  &   16.5  &  5354  &    9.0  &   0.32  \\
  6  &  -1.13  &  -2.40  &    1.4  &    6.0  &  3403  &    3.3  &   0.25  \\
  7  &  -0.75  &  -3.08  &    0.9  &    5.5  &  4038  &    3.0  &   0.16  \\
  8  &  -2.55  &  -0.38  &    0.8  &    4.4  &  3531  &    2.4  &   0.15  \\
\hline
\multicolumn{8}{l}{G094}\\
  1  &   0.15  &   0.07  &   13.6  &   57.1  &  5295  &   36.9  &   5.18  \\
  2  &  -0.89  &  -0.37  &    2.4  &   10.4  &  2837  &    6.8  &   1.04  \\
  3  &  -2.00  &   5.27  &    2.2  &   14.7  &  3499  &    9.6  &   0.93  \\
  4  &  -1.19  &   0.74  &    1.8  &    7.7  &  2579  &    5.1  &   0.76  \\
\hline
\multicolumn{8}{l}{CepA}\\
  1  &   0.00  &   0.07  &  440.9  & 1131.7  &   964  &    2.6  &  21.10  \\
  2  &  -0.15  &  -3.33  &   85.7  &   93.1  &   296  &    0.2  &   4.10  \\
\hline
\multicolumn{8}{l}{NGC7538IRS9}\\
  1  &   0.52  &   0.74  &   41.2  &   93.7  &  3434  &    4.6  &   2.79  \\
  2  &   2.81  &   0.44  &    5.9  &   27.2  &  2293  &    1.3  &   0.40  \\
  3  &   3.18  &   0.89  &    4.9  &   25.4  &  2191  &    1.2  &   0.33  \\
  4  &   3.77  &  -0.22  &    4.7  &   41.6  &  3061  &    2.0  &   0.32  \\
  5  &   1.11  &   1.77  &    3.0  &    9.5  &  1632  &    0.5  &   0.20  \\
  6  &  -0.22  &   2.14  &    2.7  &   14.0  &  2185  &    0.7  &   0.18  \\
  7  &   1.33  &   2.66  &    1.9  &    5.4  &  1397  &    0.3  &   0.13  \\
  8  &   1.92  &   2.00  &    1.8  &    6.4  &  1526  &    0.3  &   0.12  \\
  9  &  -1.63  &   2.36  &    1.6  &   14.2  &  2446  &    0.7  &   0.11  \\
\hline
\multicolumn{8}{l}{W3H2O}\\
  1  &  -5.28  &   0.30  &  451.6  & 1879.3  &  1740  &    0.0  &   0.00  \\
  2  &  -4.90  &  -0.37  &  320.0  & 1464.2  &  1553  &    0.0  &   0.00  \\
  3  &   0.89  &  -0.07  &  172.3  &  940.6  &  2150  &   13.0  &   7.31  \\
  4  &  -0.37  &   0.00  &  165.3  &  720.3  &  1704  &   10.0  &   7.02  \\
  5  &  -1.26  &   0.74  &   47.4  &  133.9  &  1061  &    1.9  &   2.01  \\
  6  &  -1.04  &   1.41  &   19.4  &   83.7  &  1074  &    1.2  &   0.82  \\
  7  &  -2.08  &   0.74  &   19.5  &   70.1  &   985  &    1.0  &   0.83  \\
\hline
\multicolumn{8}{l}{W3IRS4}\\
  1  &   0.89  &  -0.22  &   39.3  &   92.9  &  1478  &    3.2  &   3.97  \\
  2  &   5.34  &   2.74  &   19.4  &   86.6  &  1575  &    3.0  &   1.97  \\
  3  &   2.00  &   1.26  &   13.6  &   81.2  &  1586  &    2.8  &   1.38  \\
  4  &   1.70  &   1.70  &   12.7  &   50.8  &  1295  &    1.8  &   1.29  \\
  5  &  -1.48  &  -1.33  &    9.0  &   36.6  &  1159  &    0.0  &   0.00  \\
  6  &  -0.59  &  -2.00  &    7.7  &   29.0  &  1177  &    0.0  &   0.00  \\
\hline
\multicolumn{8}{l}{G108}\\
  1  &   1.26  &   0.44  &   14.8  &   33.2  &  4392  &   10.8  &   1.99  \\
  2  &  -3.69  &   2.00  &    6.7  &   14.9  &  3187  &    4.8  &   0.89  \\
  3  &  -2.22  &   3.25  &    4.9  &   11.6  &  3054  &    3.8  &   0.66  \\
\hline
\multicolumn{8}{l}{IRAS23385}\\
  1  &   0.81  &   0.00  &   18.0  &  114.0  &  6956  &   21.9  &   1.18  \\
  2  &   2.36  &   0.44  &   14.7  &   49.2  &  6232  &    9.4  &   0.96  \\
  3  &  -0.59  &   0.15  &    4.4  &   26.7  &  5301  &    5.1  &   0.29  \\
\hline
\multicolumn{8}{l}{G138}\\
  1  &  -0.30  &  -0.37  &    6.2  &   78.7  &  4912  &   11.6  &   0.89  \\
  2  &   7.24  &  11.60  &    5.4  &   12.4  &  1539  &    1.8  &   0.78  \\
  3  &   0.07  &   5.32  &    2.0  &    8.9  &  2083  &    1.3  &   0.29  \\
\hline
\multicolumn{8}{l}{G139}\\
  1  &  -0.22  &  -5.47  &   13.9  &   19.7  &  1887  &    2.4  &   1.30  \\
  2  &   0.44  &   4.66  &    6.0  &    6.4  &  1386  &    0.8  &   0.63  \\
\hline
\multicolumn{8}{l}{NGC7538IRS9}\\
  1  &   0.07  &  -0.15  & 2334.1  & 2837.7  &  1596  &   43.0  &  70.00  \\
\hline
\multicolumn{8}{l}{NGC7538S}\\
  1  &   0.96  &   1.70  &   28.1  &   82.3  &  1337  &    4.5  &   3.24  \\
  2  &  -2.51  &  -0.37  &   23.4  &   56.4  &  1459  &    3.1  &   2.70  \\
  3  &  -0.81  &   0.52  &   21.6  &   51.2  &  1502  &    2.8  &   2.49  \\
  4  &   1.48  &   1.92  &   20.6  &   46.3  &  1131  &    2.5  &   2.38  \\
  5  &  -6.95  &   2.07  &   10.5  &   11.0  &   694  &    0.6  &   1.21  \\
  6  &  -2.07  &  -5.17  &    6.1  &    5.5  &   597  &    0.3  &   0.70  \\
\end{longtable}
Notes: For the mass calculations, the fluxes for several sources are
corrected for free-free emission. For the H{\sc ii} region W3OH, we do
not calculate masses. The cores 5 and 6 in W3IRS4 are also not used
for mass calculations because they are part of an H{\sc ii} region
\citep{tieftrunk1995}. Also S87IRS1 source 2 is taken out because all
emission should be free-free \citep{kurtz1994}. For other sources the
masses are calculated from mm fluxes that are corrected for their
free-free contribution: AFGL2591 source 1 \citep{vandertak2005b}, S106
source 1 \citep{kurtz1994}, G094 source 1 \citep{skinner1993}, G139
source 1 \citep{manjarrez2012}, NGC7538IRS1 source 1
\citep{beuther2012c}.}

\begin{appendix}

\section{Simulating CORE observations}
\label{simulations}

\begin{figure*}[ht]
\includegraphics[width=0.99\textwidth]{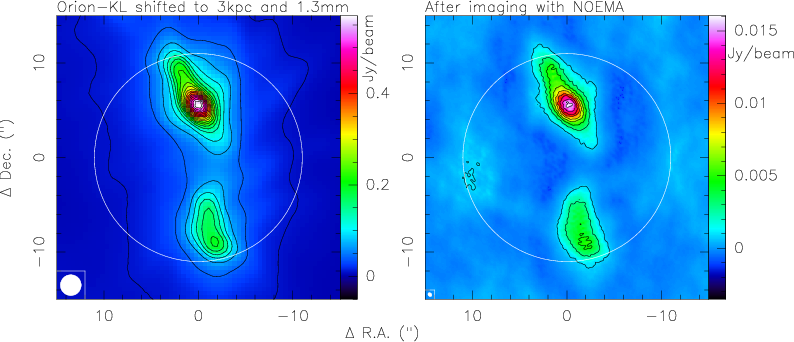}\\
\caption{Simulated CORE observations at 1.3\,mm wavelength using
  original data obtained for Orion-KL by \citet{lane2016}. For details
  of the simulations see appendix \ref{simulations}. The beam for the
  left panel (Orion shifted to 3\,kpc) is $2.19''$, whereas the
  synthesized beam of the simulated image in the right panel is
  $0.44''\times 0.34''$ (P.A.~44\,deg). The contours in the left panel
  start at 10\,mJy\,beam$^{-1}$ and continue in 30\,mJy\,beam$^{-1}$
  steps. The contours in the right image start at $3\sigma$ and
  continue in $6\sigma$ steps ($1\sigma\sim
  0.3$\,mJy\,beam$^{-1}$). The large circles are of $22''$ diameter
  marking the FWHM of the primary beam.}
\label{sim_orionkl}
\end{figure*}

\begin{figure*}[htb]
\includegraphics[width=0.99\textwidth]{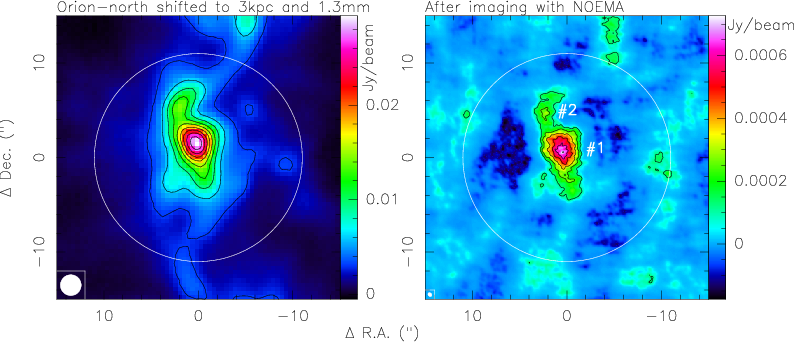}
\caption{Simulated CORE observations at 1.3\,mm wavelength using
  original data obtained for a field in the northern filament of Orion
  A by \citet{lane2016}. For details of the simulations see appendix
  \ref{simulations}. The beam for the left panel (Orion shifted to
  3\,kpc) is $2.19''$, whereas the synthesized beam of the simulated
  image in the right panel is $0.44''\times 0.34''$
  (P.A.~44\,deg). The contours in the left panel are in
  3\,mJy\,beam$^{-1}$ steps. The contours in the right image start
  are in $3\sigma$ steps ($1\sigma\sim 0.04$\,mJy\,beam$^{-1}$). The
  large circles are of $22''$ diameter marking the FWHM of the primary
  beam.}
\label{sim_north}
\end{figure*}

\begin{figure}[ht]
\includegraphics[width=0.49\textwidth]{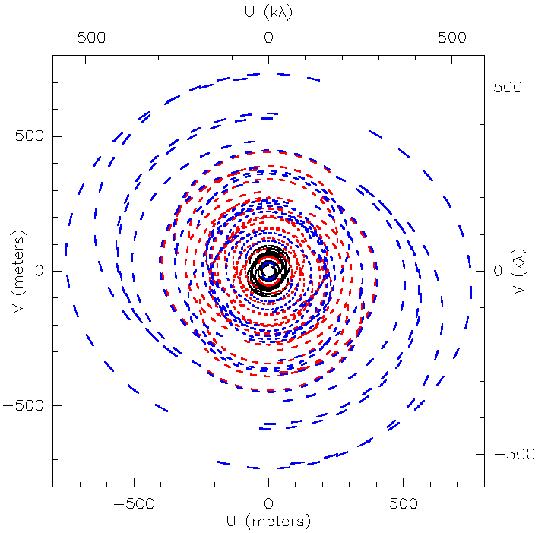}
\caption{Simulated uv-coverage.}
\label{uv_sim}
\end{figure}

To get a better quantitative understanding on how much the imaging and
spatial filtering of the interferometer affects our results, we
simulated CORE observations using a distance-scaled bolometer dataset
from the Orion molecular cloud. The original Orion 850\,$\mu$m data
are the large-scale SCUBA-2 observations of Orion A by
\citet{lane2016}.

For the simulations, several steps had to be applied to this original
dataset. (i) Since our CORE data are at 1.3\,mm wavelength, we scaled
the flux densities $S$ of the SCUBA-2 850\,$\mu$m data by a typical
spectral index at (sub)mm wavelength of $S\propto \nu^{3.5}$. (ii) We
adapted the intrinsic spatial resolution of the data from the 450\,pc
distance of Orion \citep{lane2016} to 3\,kpc (the typical distance of
the CORE sample). This implied that the flux densities decrease by
$S\propto \left(\frac{450}{3000}\right)^{2}$. (iii) The angular
resolution of the original SCUBA-2 data of $14.6''$ corresponds at the
given distance of 450\,pc to a linear resolution of 6570\,AU. To
maintain the same linear resolution at 3\,kpc distance, we changed the
pixel size of the image accordingly (from $3''$ to $0.45''$ pixel
size). (iv) Furthermore, we converted the units of the image from
mJy\,arcsec$^{-2}$ to mJy\,beam$^{-1}$ and then to K.

From the originally large Orion A image, we selected two sub-regions
as examples: (a) a very strong region (Orion-KL), and (b) a
fragmented weaker core within the northern part of the integral shape
filament (refered to as ``Orion-north'' in the remaining part of the
section). These two files are our model images that are run through
the NOEMA simulator as described below. The two model images in units
of mJy\,beam$^{-1}$ are shown in Figures \ref{sim_orionkl} and
\ref{sim_north} (left panels).

\begin{table}[htb]
\caption{Simulation parameters}
\begin{tabular}{lrr}
\hline \hline
 & Orion-KL & Orion-north \\
\hline
rms (mJy\,beam$^{-1}$) & 0.3 & 0.04 \\
rms (mK) & 49.8 & 6.6 \\ 
$3\sigma~N_{\rm{H_2}}$ sensitivity @50\,K (cm$^{-2}$) & $1.5\times 10^{23}$ & $1.9\times 10^{22}$\\  
$3\sigma~m_{\rm{H_2}}$ sensitivity @50\,K (M$_{\odot}$) & 0.1 & 0.01 \\  
missing flux (\%) & 55 & 72 \\
$S_{\rm{peak}}(\#1)$ (mJy\,beam$^{-1}$) & & 0.73 \\
$S_{\rm{int}}(\#1)$ (mJy) & & 40 \\
$S_{\rm{int}}(\#1)$-model (mJy) & & 99 \\
$S_{\rm{peak}}(\#2)$ (mJy\,beam$^{-1}$) & & 0.3 \\
$S_{\rm{int}}(\#2)$ (mJy) & & 10 \\
$S_{\rm{int}}(\#2)$-model (mJy) & & 28 \\
\hline \hline
\end{tabular}
\label{sim_par}
\end{table}

The actual NOEMA simulations were then conducted within the \textsc{gildas}
package. As a first step, a representative uv-coverage has to be
created within the ASTRO-sub-package by the command {\sf uv\_tracks}. To
emulate our simulations best, we used a setup employing the D, B and
A-configurations. In each configuration the source was visited 15
times for a length of 15\,min each with separations of 40\,min between
the visits. As target coordinates we used W3(H$_2$O) (Table
\ref{sample}). The resulting uv-coverage is shown in
Fig.~\ref{uv_sim}. 

Finally, the visibilities are produced within the \textsc{gildas} sub-package
MAPPING and the task {\sf uv\_fmodel} using the above described model
images (converted to K) and the given uv-coverage. The resulting
visibility data files can then be imaged in exactly the same way as
our original CORE data described in section \ref{obs}. The synthesized
beam of the simulated data is $0.44''\times 0.34''$ (P.A.~44\,deg).

As for the original CORE data, the rms varies corresponding to the
side-lobes produced by the strongest sources in the field.  In the
strong Orion-KL field, the rms is 0.3\,mJy\,beam$^{-1}$ whereas it is
almost an order of magnitude lower (0.04\,mJy\,beam$^{-1}$) in the
weaker Orion-north field. For details of the simulation parameters see
Table \ref{sim_par}. 

Quantitatively, we extracted the fluxes toward the two main sources
(\#1 and \#2) in Orion-north (Fig.~\ref{sim_north}, Table
\ref{sim_par}). Assuming optically thin dust emission at mm
wavelengths with an assumed dust temperature of 50\,K, we can
calculate the corresponding core masses as in the main part of the
paper (section \ref{results}). The approximate core masses of sources
\#1 and \#2 in Orion-north are then 10.6 and 3.0\,M$_{\odot}$,
respectively. In the Orion-north field itself with the lower rms, we
detect both sources well, however, with reduced integrated fluxes at
35 to 40\% of the model fluxes (Table \ref{sim_par}). Hence, the core
masses are underestimated by the same factor. The situation would be
different if these two sources were in the same field as
Orion-KL. With an rms of 0.3\,mJy\,beam$^{-1}$ in the Orion-KL
simulation, the peak flux densities of sources \#1 and \#2 of 0.73 and
0.3\,mJy\,beam$^{-1}$ are not above the $3\sigma$ thresholds.
 
If these Orion-north sources are typical for star-forming regions,
then in fields with strong main sources, sub-sources of up
$\sim$10\,M$_{\odot}$ could be difficult to detect. In contrast to
that, in fields with weaker main sources, our core mass
sensitivity should extend down to 1\,M$_{\odot}$ or even
lower. Therefore, our core-mass sensitivity within the CORE survey is
dynamic-range and source-structure limited, and depends on the
strongest source in the field. While for weaker sources like
Orion-north, we reach almost the thermal noise limit, for stronger
sources like Orion-KL, the dynamic range limit of these simulations is
$\sim$53.

It should be noted that the above derived core mass sensitivity
strongly depends on the size of the cores and hence how much we
resolve them. As shown in Table \ref{sim_par}, the point source mass
sensitivity in both simulations is much smaller at 0.1 and
0.01\,M$_{\odot}$, respectively. However, this is only valid for point
sources. If our model cores were much more compact and all flux within
a single synthesized beam, we could easily detect them. But since we
know from the original Orion observations their actual sizes, the
emission extends over many beams which reduces the mass sensitivity
for extended objects significantly.

If we now compare these simulations to the actual range of identified
core masses in the survey (Fig.~\ref{histo}), we find that most cores
have masses between $\sim$0.2 and 6\,M$_{\odot}$, below the regime
estimated from our simulations. This difference can on the one hand be
attributed to the missing flux implying that the core masses
themselves are underestimated. But on the other hand, the source
structure may also be different in our target regions compared to the
Orion data used in these simulations. \citet{beuther2015a} have
compared three different starless regions from the low-mass B68 to the
intermediate-mass IRDC\,19175 and the high-mass IRDC\,18310-4
region. They found that the average densities between these three
regions varied between $10^4$\,cm$^{-3}$ for the low-mass case to
$10^6$\,cm$^{-3}$ for the high-mass region. While the Orion-north
region within the integral-shape filament forms rather low- to
intermediate-mass stars (in contrast to the high-mass region
Orion-KL), our sample is selected to form high-mass stars. Hence,
higher densities and more compact structures are expected for the CORE
sample. As outlined above, such compact structures make detections of
lower-mass objects easier than if the emission is distributed over
larger areas. Such structural differences may partly explain the
differences between the simulated mass sensitivities and the observed
ones.

\section{Individual continuum images}
\label{individual_images}

\begin{figure*}[ht]
\includegraphics[width=0.49\textwidth]{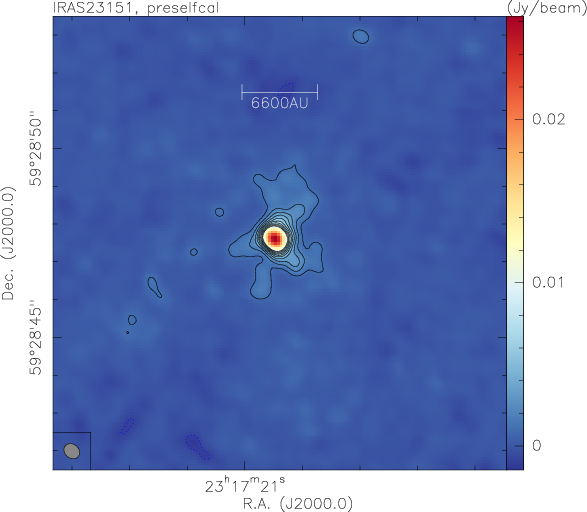}
\includegraphics[width=0.49\textwidth]{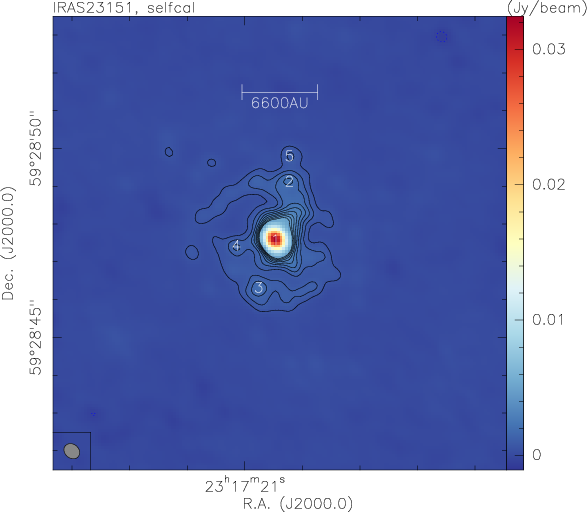}\\
\includegraphics[width=0.49\textwidth]{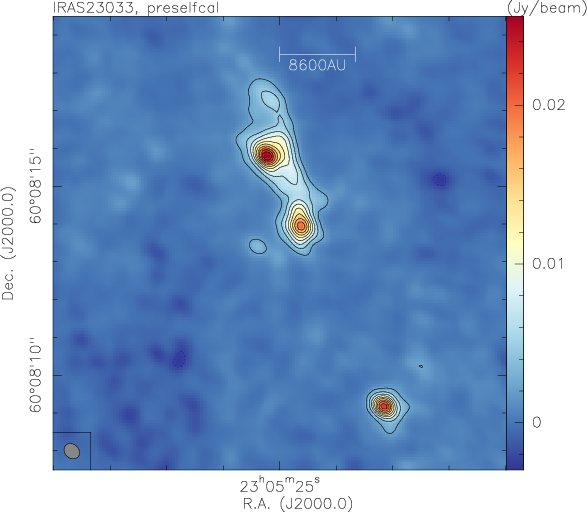}
\includegraphics[width=0.49\textwidth]{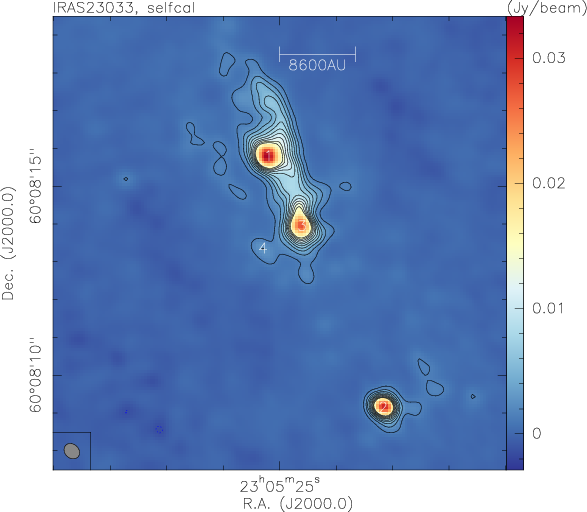}\\
\includegraphics[width=0.49\textwidth]{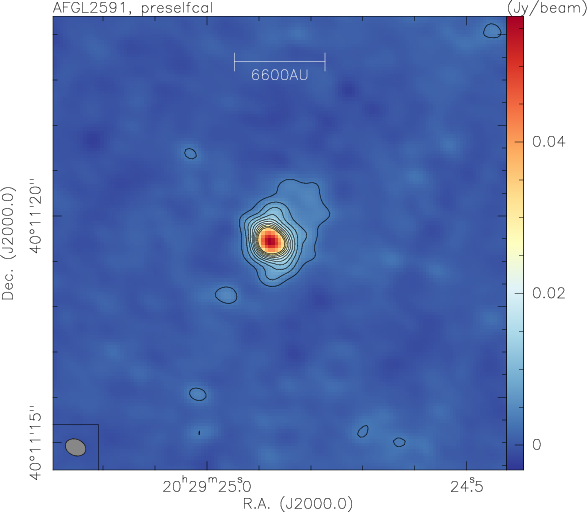}
\includegraphics[width=0.49\textwidth]{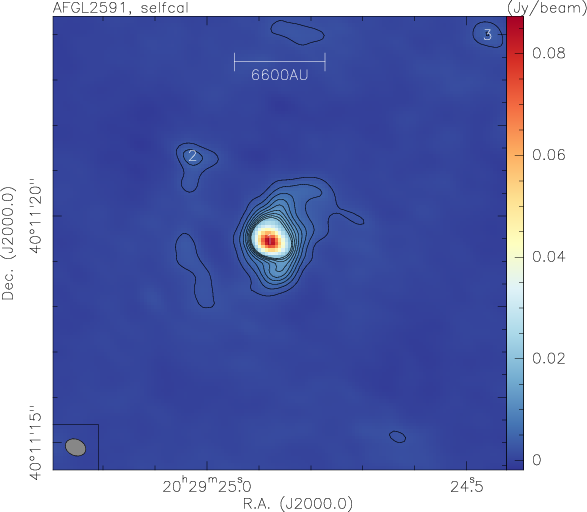}
\caption{1.37\,mm continuum data for CORE sources. The left panels
  always show the data without applying self-calibration, and the right
  panels show them after applying self-calibration. The contours are
  always in 5$\sigma$ steps (see table \ref{para}). The right panels
  mark the cores identified with clumpfind.}
\end{figure*}

\begin{figure*}[ht]
\includegraphics[width=0.49\textwidth]{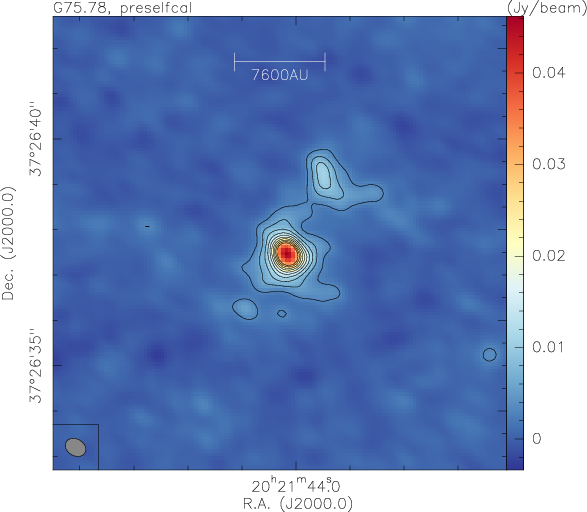}
\includegraphics[width=0.49\textwidth]{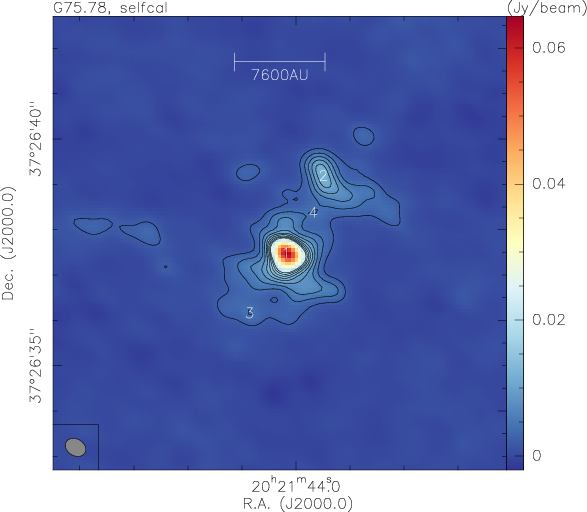}\\
\includegraphics[width=0.49\textwidth]{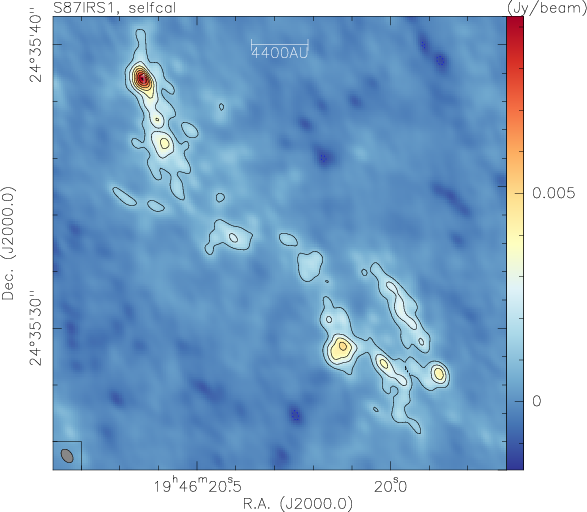}
\includegraphics[width=0.49\textwidth]{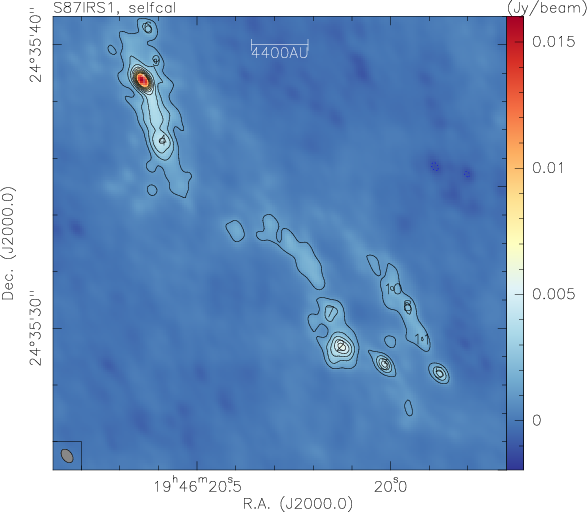}\\
\includegraphics[width=0.49\textwidth]{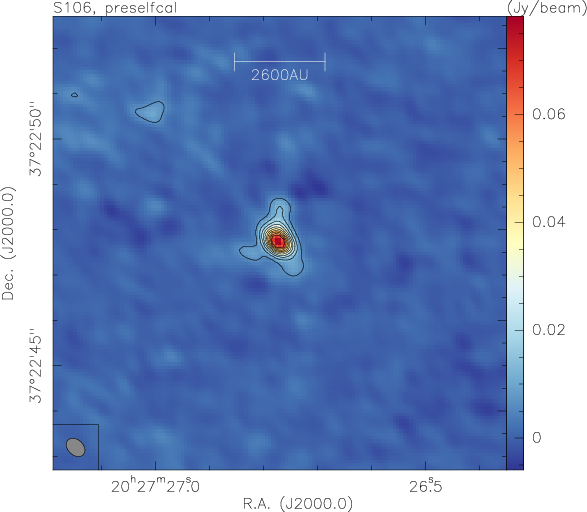}
\includegraphics[width=0.49\textwidth]{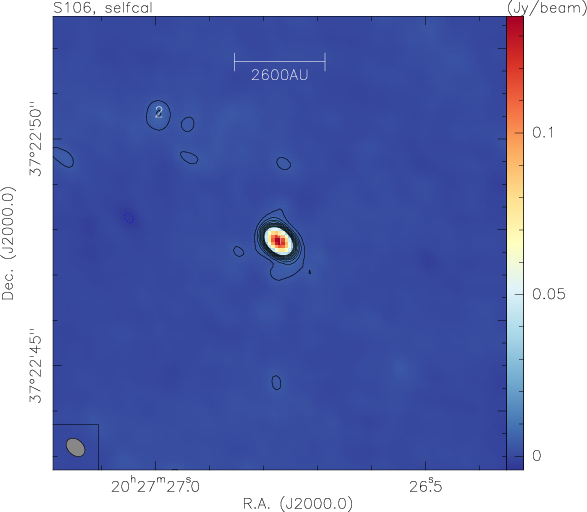}
\caption{1.37\,mm continuum data for CORE sources. The left panels
  always show the data without applying self-calibration, and the right
  panels show them after applying self-calibration. The contours are
  always in 5$\sigma$ steps (see table \ref{para}). The right panels
  mark the cores identified with clumpfind.}
\end{figure*}

\begin{figure*}[ht]
\includegraphics[width=0.49\textwidth]{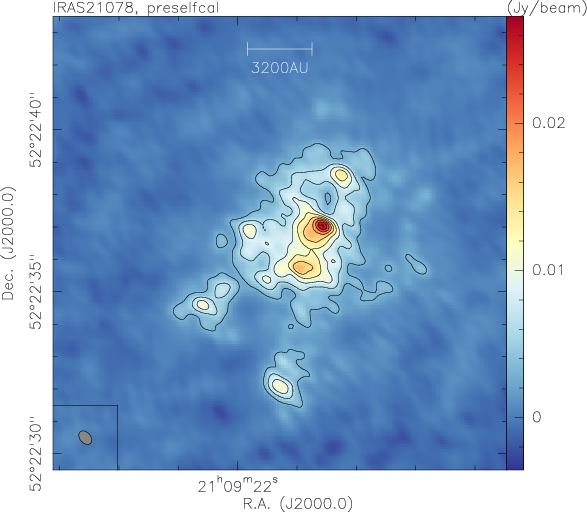}
\includegraphics[width=0.49\textwidth]{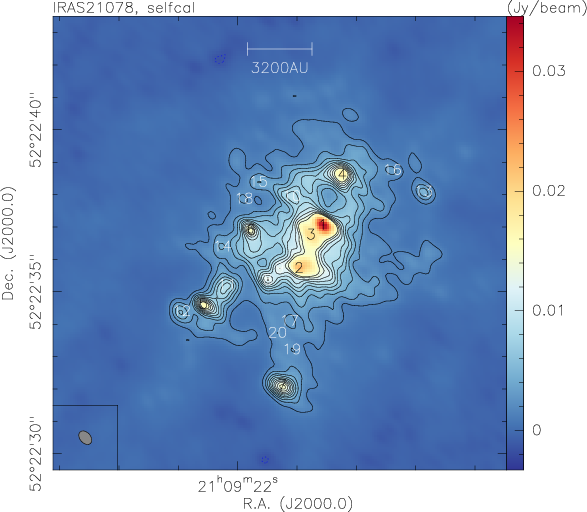}\\
\includegraphics[width=0.49\textwidth]{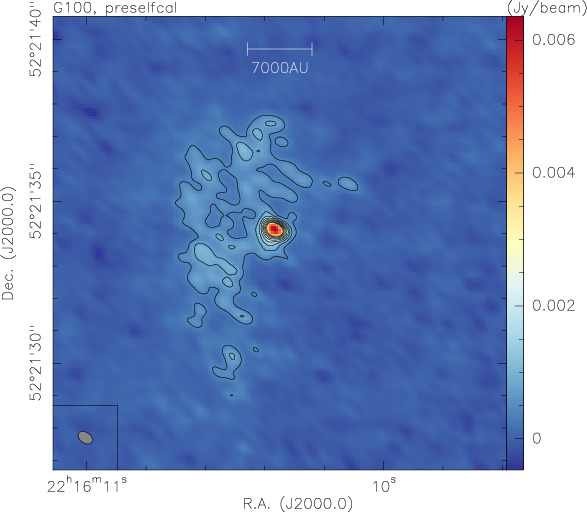}
\includegraphics[width=0.49\textwidth]{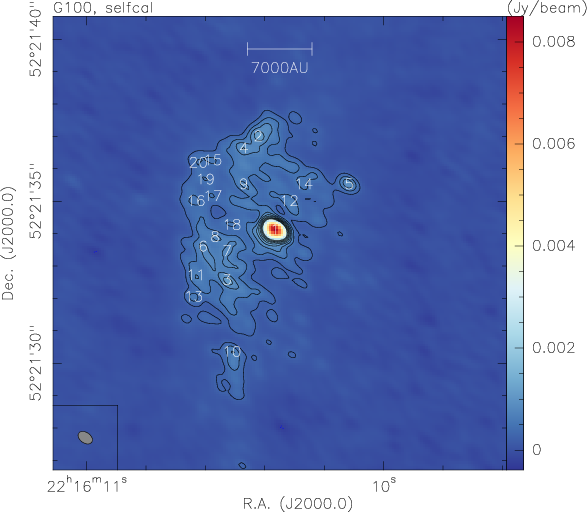}\\
\includegraphics[width=0.49\textwidth]{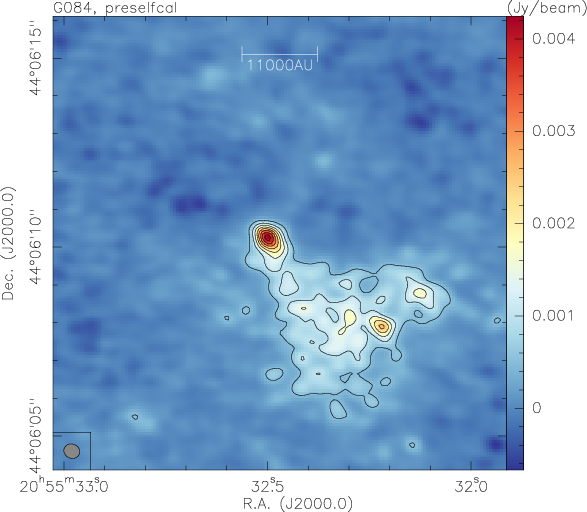}
\includegraphics[width=0.49\textwidth]{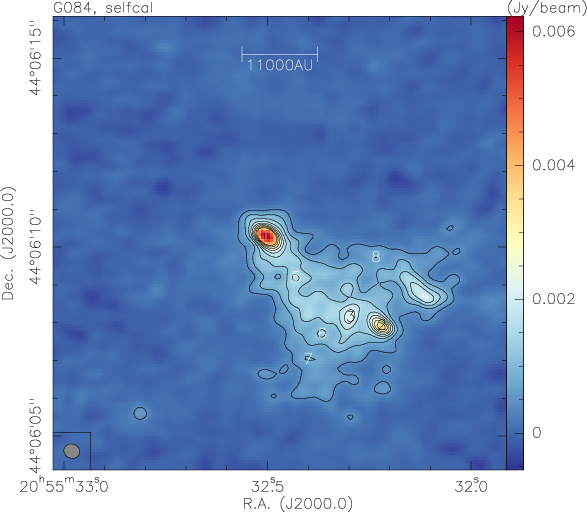}
\caption{1.37\,mm continuum data for CORE sources. The left panels
  always show the data without applying self-calibration, and the right
  panels show them after applying self-calibration. The contours are
  always in 5$\sigma$ steps (see table \ref{para}). The right panels
  mark the cores identified with clumpfind.}
\end{figure*}

\begin{figure*}[ht]
\includegraphics[width=0.49\textwidth]{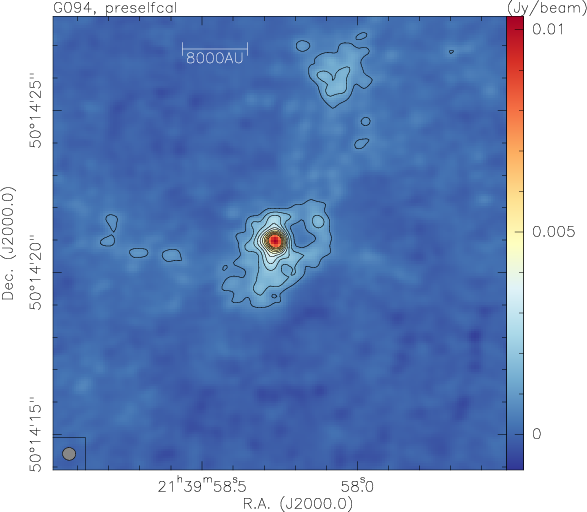}
\includegraphics[width=0.49\textwidth]{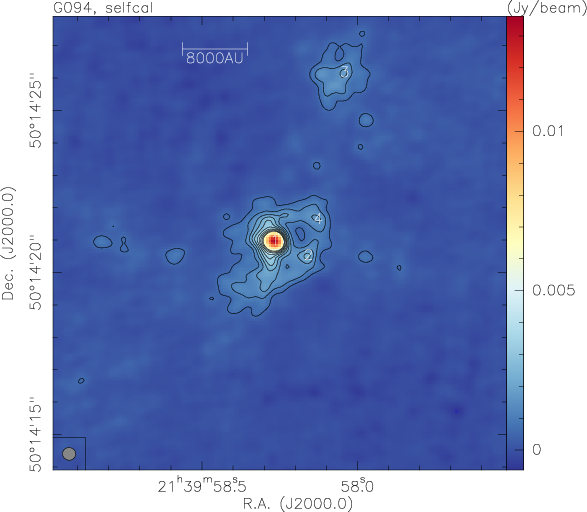}\\
\includegraphics[width=0.49\textwidth]{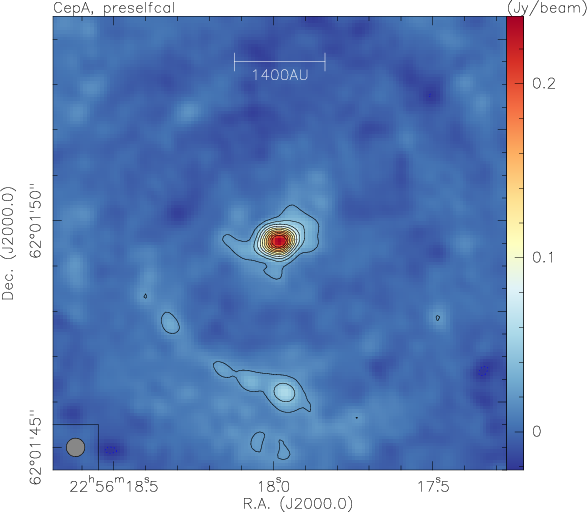}
\includegraphics[width=0.49\textwidth]{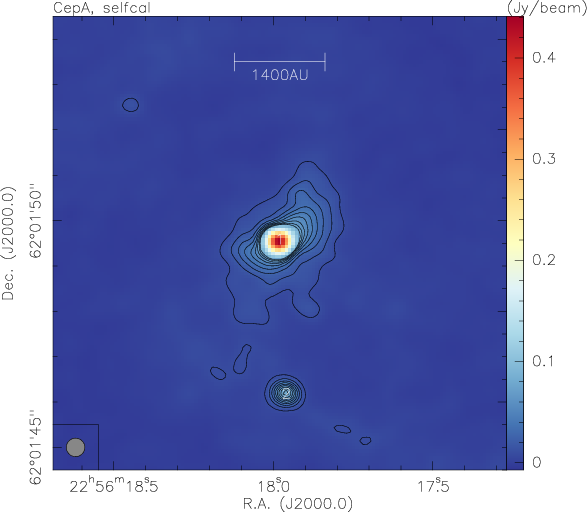}\\
\includegraphics[width=0.49\textwidth]{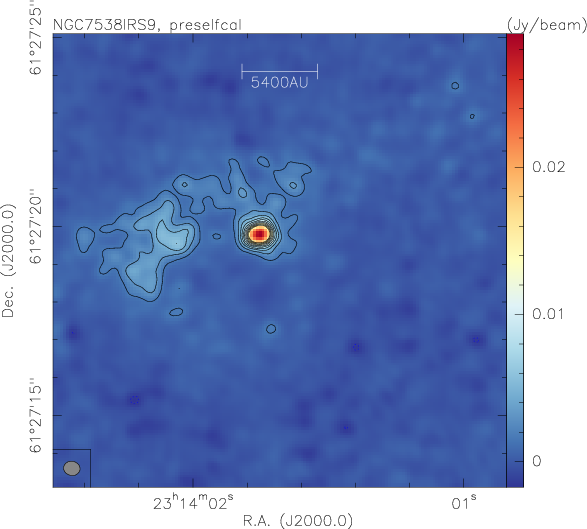}
\includegraphics[width=0.49\textwidth]{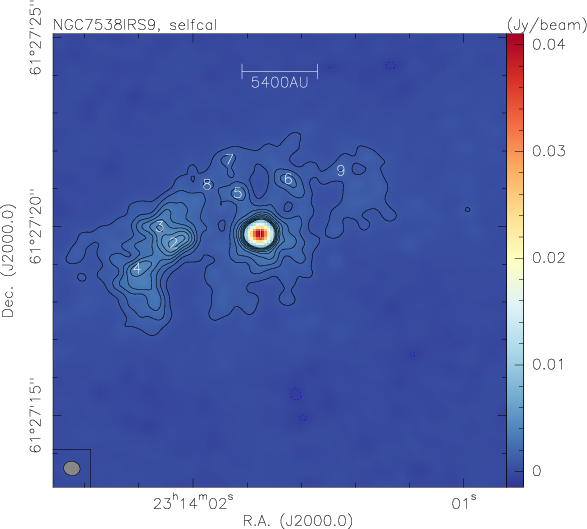}
\caption{1.37\,mm continuum data for CORE sources. The left panels
  always show the data without applying self-calibration, and the right
  panels show them after applying self-calibration. The contours are
  always in 5$\sigma$ steps (see table \ref{para}). The right panels
  mark the cores identified with clumpfind.}
\end{figure*}

\begin{figure*}[ht]
\includegraphics[width=0.49\textwidth]{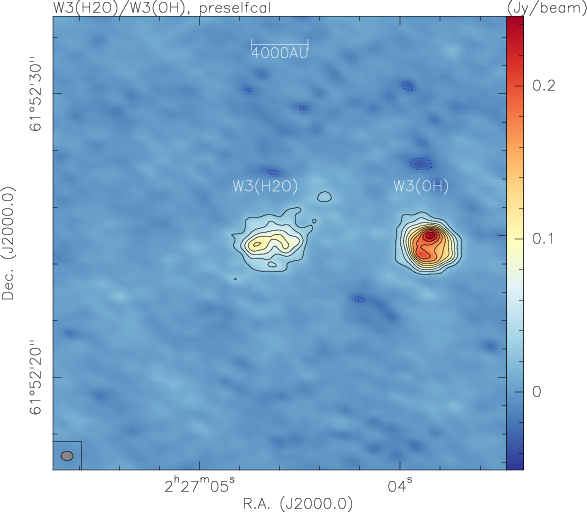}
\includegraphics[width=0.49\textwidth]{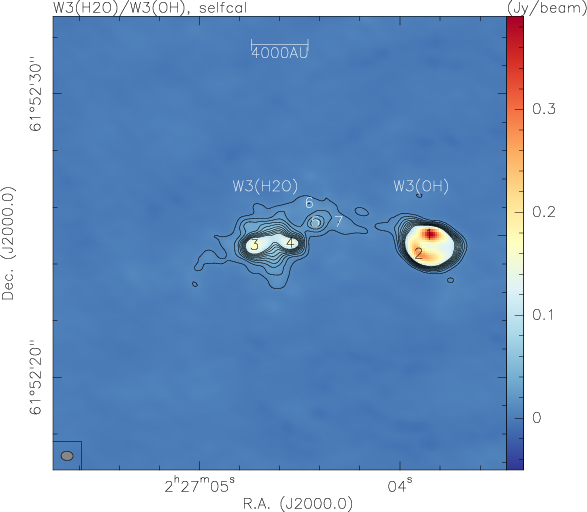}\\
\includegraphics[width=0.49\textwidth]{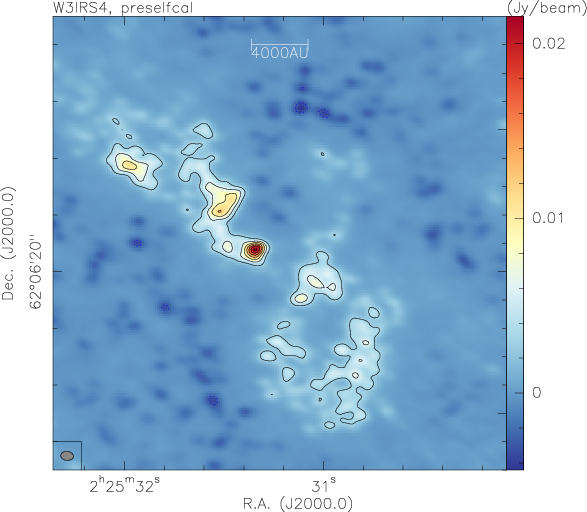}
\includegraphics[width=0.49\textwidth]{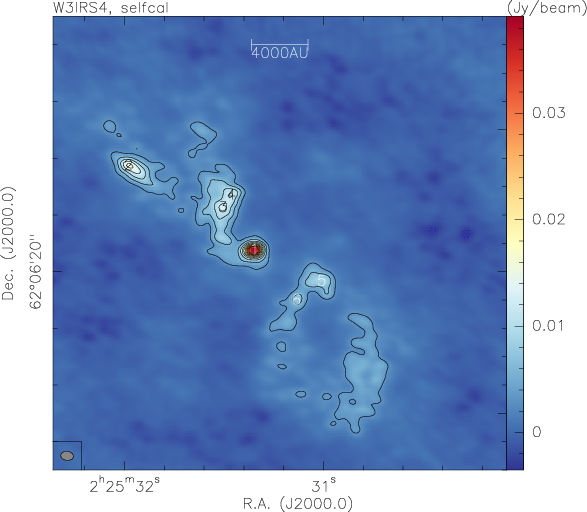}\\
\includegraphics[width=0.49\textwidth]{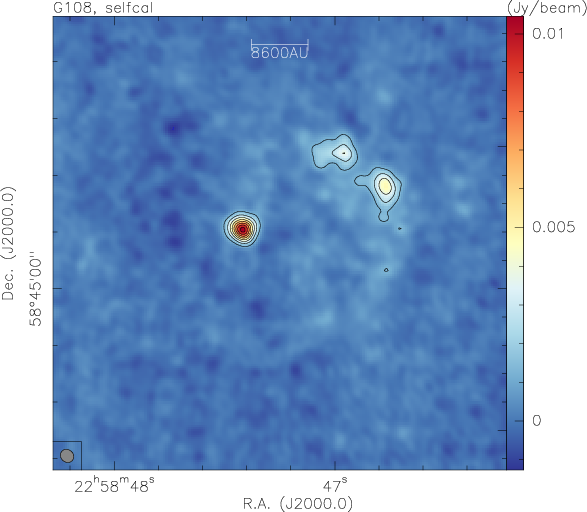}
\includegraphics[width=0.49\textwidth]{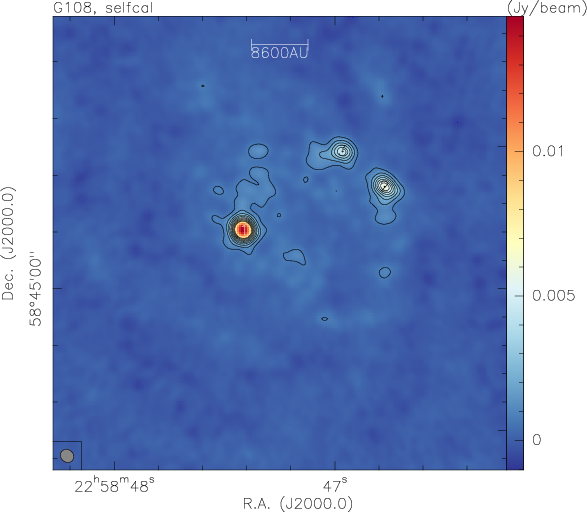}
\caption{1.37\,mm continuum data for CORE sources. The left panels
  always show the data without applying self-calibration, and the right
  panels show them after applying self-calibration. The contours are
  always in 5$\sigma$ steps (see table \ref{para}). The right panels
  mark the cores identified with clumpfind.}
\end{figure*}

\begin{figure*}[ht]
\includegraphics[width=0.49\textwidth]{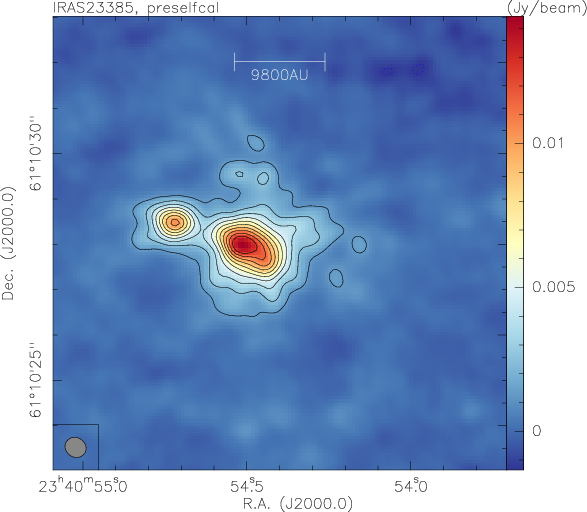}
\includegraphics[width=0.49\textwidth]{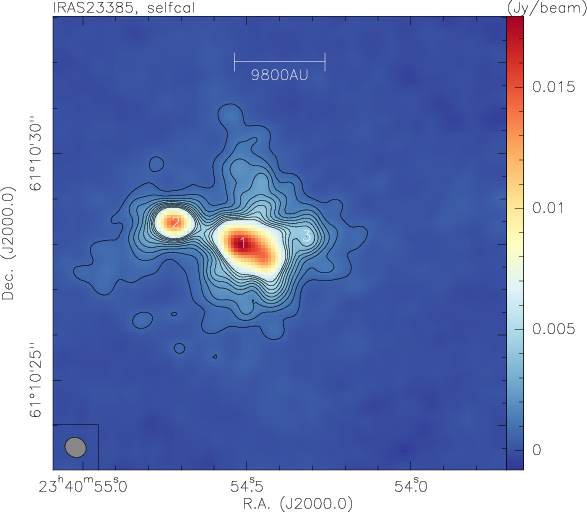}\\
\includegraphics[width=0.49\textwidth]{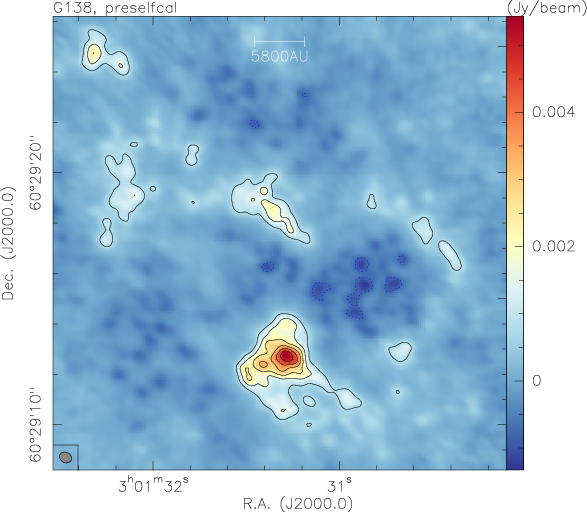}
\includegraphics[width=0.49\textwidth]{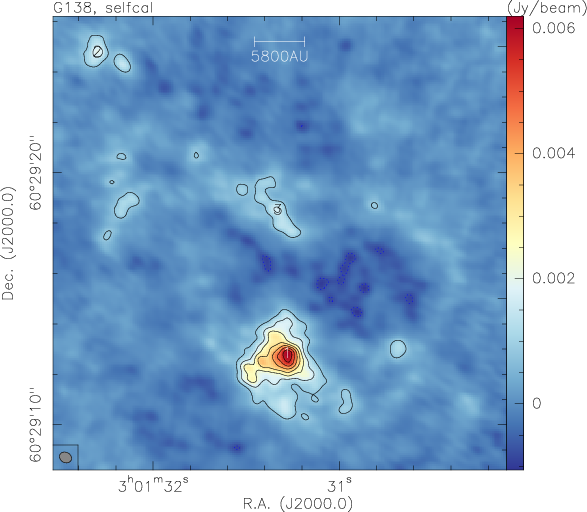}\\
\includegraphics[width=0.49\textwidth]{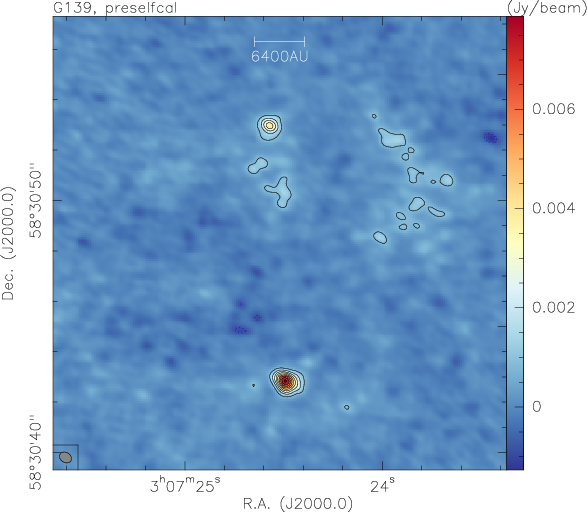}
\includegraphics[width=0.49\textwidth]{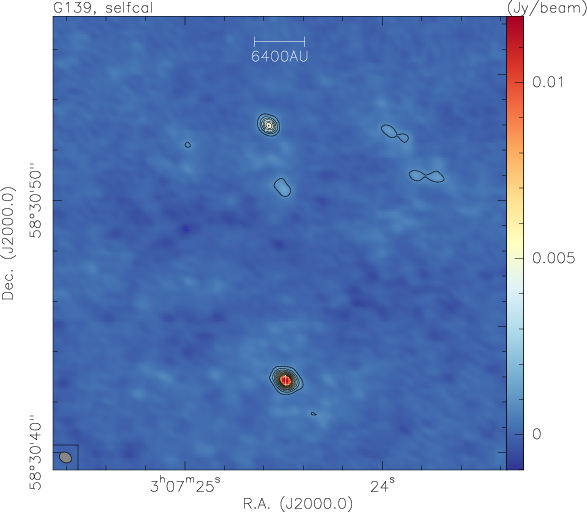}
\caption{1.37\,mm continuum data for CORE sources. The left panels
  always show the data without applying self-calibration, and the right
  panels show them after applying self-calibration. The contours are
  always in 5$\sigma$ steps (see table \ref{para}). The right panels
  mark the cores identified with clumpfind.}
\end{figure*}

\begin{figure*}[ht]
\includegraphics[width=0.49\textwidth]{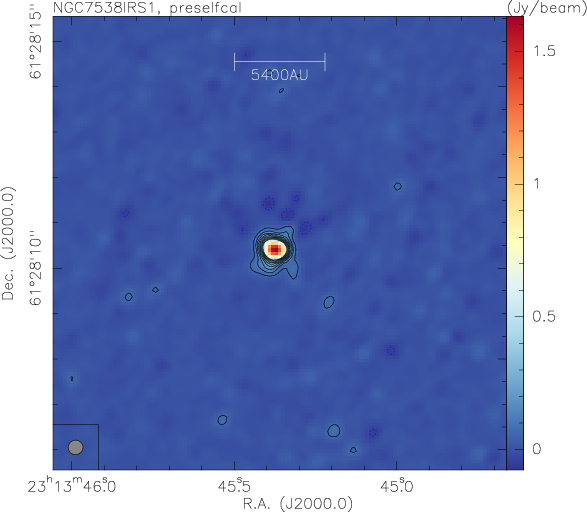}
\includegraphics[width=0.49\textwidth]{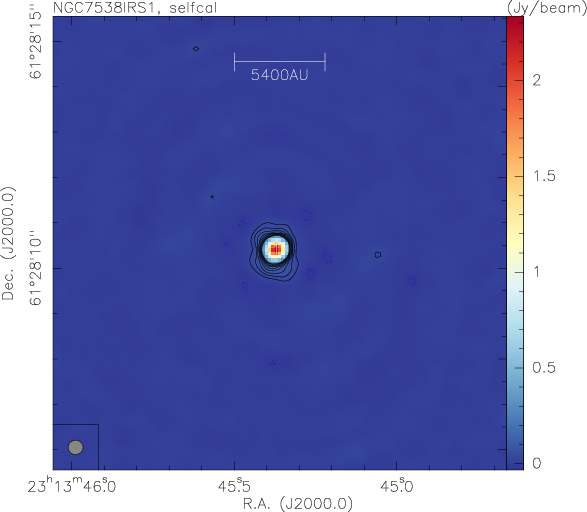}\\
\includegraphics[width=0.49\textwidth]{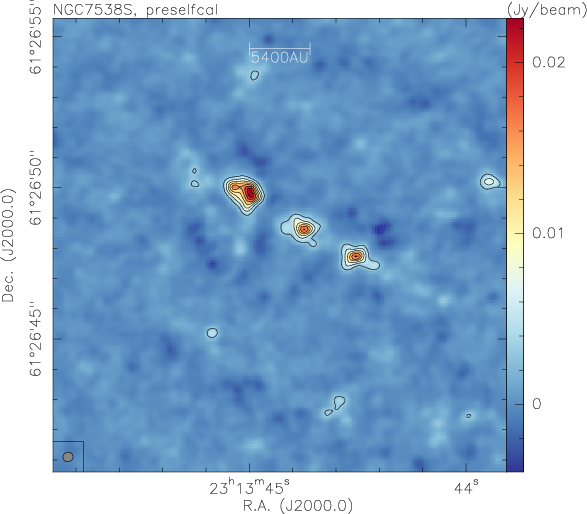}
\includegraphics[width=0.49\textwidth]{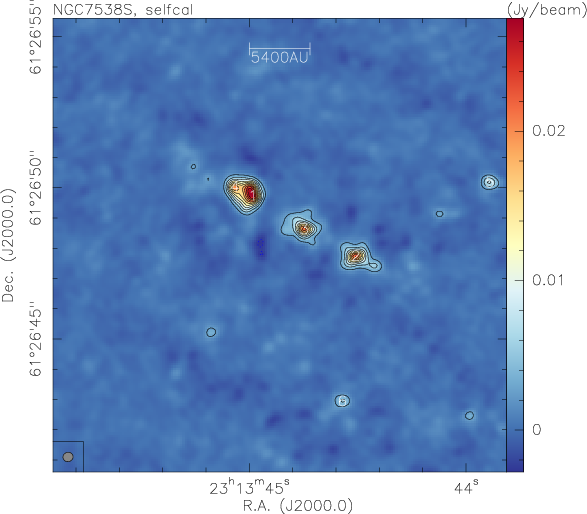}\
\caption{1.37\,mm continuum data for two CORE pilot sources. The left
  panels always show the data without applying self-calibration, and
  the right panels show them after applying self-calibration. The
  contours are always in 5$\sigma$ steps (see table \ref{para}). The
  right panels mark the cores identified with clumpfind.}
\end{figure*}

\clearpage

\end{appendix}

\end{document}